\documentclass{amsart}
\usepackage{amssymb}

\usepackage{graphicx}

\input{tcilatex}
\input tcilatex

\begin{document}
\title[branching diffusion and selection]{A branching diffusion model of selection: from the neutral Wright-Fisher
case to the one including mutations}
\author{Thierry E. Huillet}
\address{Laboratoire de Physique Th\'{e}orique et Mod\'{e}lisation\\
CNRS-UMR 8089 et Universit\'{e} de Cergy-Pontoise\\
2 Avenue Adolphe Chauvin, F-95302, Cergy-Pontoise, France\\
E-mail: Thierry.Huillet@u-cergy.fr}
\maketitle

\begin{abstract}
We consider\textbf{\ }diffusion processes $x_{t}$ on the unit interval.
Doob-transformation techniques consist of a selection of $x_{t}-$paths
procedure. The law of the transformed process is the one of a branching
diffusion system of particles, each diffusing like a new process $\widetilde{%
x}_{t},$ superposing an additional drift to the one of $x_{t}$. Killing
and/or branching of $\widetilde{x}_{t}-$particles occur at some
space-dependent rate $\lambda $. For this transformed process, so in the
class of branching diffusions, the question arises as to whether the
particle system is sub-critical, critical or super-critical. In the first
two cases, extinction occurs with probability one.

We apply this circle of ideas to diffusion processes arising in population
genetics. In this setup, the process $x_{t}$ is a Wright-Fisher (WF)
diffusion, either neutral or with mutations.

We study a particular Doob transform which is based on the exponential
function in the usual fitness parameter $\sigma $. We have in mind that this
is an alternative way to introduce selection or fitness in both WF-like
diffusions, leading to branching diffusion models ideas. For this
Doob-transform model of fitness, the usual selection drift $\sigma x\left(
1-x\right) $ should be superposed to the one of $x_{t}$ to form $\widetilde{x%
}_{t}$ which is the process that can branch, binarily.

In the first neutral case, there is a trade-off between branching events
giving birth to new particles and absorption at the boundaries, killing the
particles. Under our assumptions, the branching diffusion process gets
eventually globally extinct in finite time with exponential tails.

In the second case with mutations, there is a trade-off between killing
events removing some particles from the system and reflection at the
boundaries where the particles survive. This branching diffusion process
also gets eventually globally extinct but in very long finite time with
power-law tails.

Our approach relies on the spectral expansion of the transition probability
kernels of both $x_{t}$ and $\widetilde{x}_{t}$.\newline

\textbf{Running title:} Branching diffusion model of selection.\newline

\newline

\end{abstract}

\section{Introduction}

We consider diffusion processes on the unit interval with in mind a series
of elementary stochastic models arising chiefly in population dynamics.
Special emphasis is put on Doob-transformation techniques of the diffusion
processes under concern.

Most of the manuscript's content focuses on the specific Wright-Fisher (WF)
diffusion model and some of its variations, describing the evolution of one
two-locus colony undergoing random mating, possibly under the additional
actions of mutation and selection. These models found their way over the
last sixty years, chiefly in mathematical population genetics. We refer to
the general monographs \cite{Crow}, \cite{Mar}, \cite{Ew}, \cite{Durr} and 
\cite{Gi}.

We now describe the content of this work in some more details.\newline

Section $2$ is devoted to generalities on one-dimensional diffusions on the
unit interval $\left[ 0,1\right] $, say $\left( x_{t};t\geq 0\right) $.
Special emphasis is put on the Kolmogorov backward and forward equations,
while stressing the crucial role played by the boundaries in such
one-dimensional diffusion problems. Some questions such as the meaning of
speed and scale functions, existence of an invariant measure, random time
change... are addressed in the light of the Feller classification of
boundaries. When the boundaries are absorbing, the important problem of
evaluating additive functionals $\alpha $ along sample paths is then briefly
discussed, emphasizing the prominent role played by the Green function of
the model.

So far, we have dealt with a given diffusion process $x_{t},$ and recalled
the various ingredients for computing the expectations of various quantities
of interest, summing up over the history of its paths. In this setup, there
is no distinction among paths with different destinations, nor did we allow
for annihilation or creation of paths inside the domain. The Doob transform
of paths is an invitation to do so. This important class of transformations
is a particular instance of a more general construction based on
multiplicative functionals. We fix the background.

Roughly speaking, the Doob transformation of paths procedure allows to
select sample paths $x\rightarrow y$ within any laps of time $t,$ favoring
large values of the ratio $\alpha \left( y\right) /\alpha \left( x\right) $
for some specific functional $\alpha >0$ that fixes the selection problem
under study$.$ The process solving this selection of paths procedure belongs
to a class of branching diffusion processes, where independent particles
diffusing like a new process $\widetilde{x}_{t}$ inside the interval are
allowed to duplicate would the visited region of the state-space fulfill the
selection of paths criterion or to die, if not. In the process, advantageous
regions of the state-space are reinforced while unfavorable ones are left
unexplored which is a reasonable physical way to look at selection of paths.
The new process $\widetilde{x}_{t}$ alluded to is obtained from $x_{t}$ just
after superposing an additional suitable drift to the latter process. An
important parameter is the state-dependent rate $\lambda $ at which killing
and/or branching occur$.$ Depending on $\lambda $ and on the type of
boundaries which $\left\{ 0,1\right\} $ are to $\widetilde{x}_{t},$ the full
transformed process can have two stopping times: the time to absorption at
the boundaries and the killing time inside the domain. Besides, there is or
not an opportunity that the $\widetilde{x}_{t}-$particles duplicate, leading
or not to a daughters particle system evolving independently starting from
where their mother particle died. The killing/branching issues depend on the
sign of $\lambda .$

It turns out that the same diffusion methods used in the previous discussion
on simple diffusions apply to the transformed processes obtained after the
induced change of measure. We develop this circle of ideas.\newline

We next apply these general ideas to diffusion processes arising in
population genetics.

In Section $3$ we start recalling that WF diffusion models with various
drifts are continuous space-time models which can be obtained as scaling
limits of a biased discrete Galton-Watson model with a conservative number
of offsprings over the generations. Sections $4$ and $5$ are devoted to a
detailed study of both the neutral WF diffusion process (WFN), the WF
diffusion with selection (WFS), the WF diffusion with mutations (WFM) and
the WF diffusion with mutations and selection (WFMS) respectively.\newline

In this context, our suggestion is the following one: we can view the action
of selection (or fitness) on the evolution of the allele frequency
distribution, either neutral or with mutations, as a functional deformation
of the sample paths of the original process, say $x_{t},$ favoring initial
values $x_{0}=x$ with small $\alpha \left( x\right) $ and terminal values $%
x_{t}=y$ with large $\alpha \left( y\right) ,$ for each $t.$ In our
construction, $\alpha \left( x\right) =e^{\sigma x}$, $\sigma >0$, is the
chosen exponential fitness functional. Stated differently and more
precisely, if $p\left( x;t,y\right) $ is the transition probability density
of $x_{t}$ (either WFN or WFM), our model of the action of fitness is 
\begin{equation*}
p\left( x;t,y\right) \overset{\text{fitness}}{\rightarrow }\overline{p}%
\left( x;t,y\right) =\frac{e^{\sigma y}}{e^{\sigma x}}p\left( x;t,y\right) .
\end{equation*}
With this choice of $\alpha $, the modification consists of selecting those
paths $x\rightarrow y$ of $x_{t}$ for which $e^{\sigma \left( y-x\right) }$
is large. As a result of this transformation of paths, the usual positive
selection drift $\sigma x\left( 1-x\right) $ has to be superposed to the one
of $x_{t}$ to form the new process $\widetilde{x}_{t}$, but our functional
definition of fitness also generates an additional branching multiplicative
term, translating that a particle system pops in: The resulting transformed
process is a (binary) branching diffusion of WF diffusions $\widetilde{x}%
_{t} $. We may call the obtained processes the branching neutral
Wright-Fisher process and the branching Wright-Fisher process with
mutations. This point of view seems to be new, to the best of the author's
knowledge.

Because the spectral representation of both transition probability densities
of WFN or WFM are known explicitly from the works of Crow and Kimura (see 
\cite{Kim1}, \cite{Crow} and \cite{CK}), some easy consequences on the
spectral structures of the branching transformed processes are available.
For instance, it is possible to decide whether the BD process is
sub-critical, critical or super-critical in the sense of (\cite{AH1} and 
\cite{AH2}).

In Section $6$, we therefore give a detailed study of the binary branching
diffusion process obtained while using the Doob transform $\alpha \left(
x\right) =e^{\sigma x}$ when the starting point process is a WFN diffusion
process. We end up with a branching particle system, each diffusing
according to the WF model with a selection drift, but branching at a bounded
rate $b>0$. In this setup, the particles cannot get killed, rather they are
allowed either to survive or to split: the transformed process is a pure
binary branching diffusion. For this super-critical binary branching
diffusion process, there is a trade-off between branching events giving
birth to new particles and absorption at the boundaries, killing the
particles. Thanks to the spectral representation of the WFN process, this
problem is amenable to the results obtained in (\cite{AH1} and \cite{AH2}).
Under our assumptions, the branching diffusion process turns out to be
globally sub-critical: the branching diffusion process gets eventually
globally extinct in finite exponential time. This requires the computation
of the ground states associated with the smallest nonnegative eigenvalue of
the infinitesimal generator of the transformed process which are here shown
to be explicit. In particular, the expression of the quasi-stationary
distribution of the particle system can be obtained in closed-form.

In Section $7$, we study the binary branching diffusion process obtained
while using the same Doob transform, when the starting point process is now
a WF diffusion process with mutations, assuming reflecting boundaries. We
end up in a branching particle system, each diffusing according to the WF
model with a mutation and selection drift, but branching at quadratic rate $%
\lambda ,$ which is bounded from below and above. Although the particles are
still allowed to split, they can now also get killed at the branching times:
the transformed process is again a binary branching diffusion but with
killing now allowed. In this setup, there is a competition between
branching/ killing events and reflection at the boundaries where the
particles survive. This problem is also amenable to the results obtained in (%
\cite{AH1} and \cite{AH2}) and we end up now in a globally critical
branching particle system, each diffusing according to the WF model with a
mutation and selection drift. This branching diffusion process turns out to
be globally critical: it also gets eventually globally extinct but now in
long finite time, with power-law tails.

\section{Diffusion processes on the unit interval and Doob transforms}

We start with generalities on one-dimensional diffusions with the WF model
and its relatives in mind. For more technical details, we refer to \cite{Dyn}%
, \cite{EK}, \cite{Kar} and \cite{Man}. We also introduce Doob transforms as
particular instances of the modification of the original diffusion process
through a multiplicative functional.

\subsection{One-dimensional diffusions on the interval $\left[ 0,1\right] $}

Let $\left( w_{t};t\geq 0\right) $ be a standard one-dimensional Brownian
(Wiener) motion. We consider a $1-$dimensional It\^{o} diffusion driven by $%
\left( w_{t};t\geq 0\right) $ on the interval say $\left[ 0,1\right] $, see 
\cite{Ito}. We assume it has locally Lipschitz continuous drift $f\left(
x\right) $ and local standard deviation (volatility) $g\left( x\right) $,
namely we consider the stochastic differential equation (SDE)$:$%
\begin{equation}
dx_{t}=f\left( x_{t}\right) dt+g\left( x_{t}\right) dw_{t}\text{, }%
x_{0}=x\in I:=\left( 0,1\right) .  \label{1}
\end{equation}
The condition on $f\left( x\right) $ and $g\left( x\right) $ guarantees in
particular that there is no point $x_{*}$ in $I$ for which $\left| f\left(
x\right) \right| $ or $\left| g\left( x\right) \right| $ would blow up and
diverge as $\left| x-x_{*}\right| \rightarrow 0$.

The Kolmogorov backward infinitesimal generator of (\ref{1}) is $G=f\left(
x\right) \partial _{x}+\frac{1}{2}g^{2}\left( x\right) \partial _{x}^{2}$.
As a result, for all suitable $\psi $ in the domain of the operator $%
S_{t}:=e^{tG}$, $u:=u\left( x,t\right) =\mathbf{E}\psi \left( x_{t\wedge
\tau _{x}}\right) $ satisfies the Kolmogorov backward equation (KBE) 
\begin{equation*}
\Bbb{\partial }_{t}u=G\left( u\right) \text{; }u\left( x,0\right) =\psi
\left( x\right) .
\end{equation*}
In the definition of the mathematical expectation $u$, we have $t\wedge \tau
_{x}:=\inf \left( t,\tau _{x}\right) $ where $\tau _{x}$ indicates a random
time at which the process should possibly be stopped (absorbed), given the
process was started in $x$. The description of this (adapted) absorption
time is governed by the type of boundaries which $\partial I:=\left\{
0,1\right\} $ are to $\left( x_{t};t\geq 0\right) .$ A classification of the
boundaries exists, due to Feller (see \cite{Kar} pp. $226$): they can be
either accessible (namely exit or regular), or inaccessible (namely entrance
or natural).

\subsection{Natural coordinate, scale, speed measure, time change}

For such Markovian diffusions, it is interesting to consider the $G-$%
harmonic coordinate $\varphi \in C^{2}$ belonging to the kernel of $G,$ i.e.
satisfying $G\left( \varphi \right) =0.$ For $\varphi $ and its derivative $%
\varphi ^{\prime }:=d\varphi /dy$, with $\left( x_{0},y_{0}\right) \in
\left( 0,1\right) $, one finds 
\begin{eqnarray*}
\varphi ^{\prime }\left( y\right) &=&\varphi ^{\prime }\left( y_{0}\right)
e^{-2\int_{y_{0}}^{y}\frac{f\left( z\right) }{g^{2}\left( z\right) }dz} \\
\varphi \left( x\right) &=&\varphi \left( x_{0}\right) +\varphi ^{\prime
}\left( y_{0}\right) \int_{x_{0}}^{x}e^{-2\int_{y_{0}}^{y}\frac{f\left(
z\right) }{g^{2}\left( z\right) }dz}dy.
\end{eqnarray*}
One should choose a version of $\varphi $ satisfying $\varphi ^{\prime
}\left( y\right) >0$, $y\in I.$ The function $\varphi $ kills the drift $f$
of $\left( x_{t};t\geq 0\right) $ in the sense that, considering the change
of variable $y_{t}=\varphi \left( x_{t}\right) ,$%
\begin{equation*}
dy_{t}=\left( \varphi ^{\prime }g\right) \left( \varphi ^{-1}\left(
y_{t}\right) \right) dw_{t}\text{, }y_{0}=\varphi \left( x\right) .
\end{equation*}
The drift-less diffusion $\left( y_{t};t\geq 0\right) $ is often termed the
diffusion in natural coordinates with state-space $\left[ \varphi \left(
0\right) ,\varphi \left( 1\right) \right] =:\varphi \left( I\right) $. Its
volatility is $\widetilde{g}\left( y\right) :=\left( \varphi ^{\prime
}g\right) \left( \varphi ^{-1}\left( y\right) \right) .$ The function $%
\varphi $ is often called the scale function.

Whenever $\varphi \left( 0\right) >-\infty $ and $\varphi \left( 1\right)
<+\infty $, one can choose the integration constants defining $\varphi
\left( x\right) $ so that 
\begin{equation*}
\varphi \left( x\right) =\frac{\int_{0}^{x}e^{-2\int_{0}^{y}\frac{f\left(
z\right) }{g^{2}\left( z\right) }dz}dy}{\int_{0}^{1}e^{-2\int_{0}^{y}\frac{%
f\left( z\right) }{g^{2}\left( z\right) }dz}dy},
\end{equation*}
with $\varphi \left( 0\right) =0$ and $\varphi \left( 1\right) =1.$ In this
case, the state-space of $\left( y_{t};t\geq 0\right) $ is again $\left[
0,1\right] ,$ the same as for $\left( x_{t};t\geq 0\right) .$

Finally, considering the random time change $t\rightarrow \theta _{t}$ with
inverse: $\theta \rightarrow t_{\theta }$ defined by $\theta _{t_{\theta
}}=\theta $ and 
\begin{equation*}
\theta =\int_{0}^{t_{\theta }}\widetilde{g}^{2}\left( y_{s}\right) ds,
\end{equation*}
the novel diffusion $\left( w_{\theta }:=y_{t_{\theta }};\theta \geq
0\right) $ is easily checked to be identical in law to a standard Brownian
motion on $\varphi \left( I\right) $. The random time $t_{\theta }$ can be
expressed as 
\begin{equation*}
t_{\theta }=\int_{0}^{\theta }m\left( \varphi ^{-1}\left( w_{\tau }\right)
\right) \left( \varphi ^{-1}\right) ^{\prime }\left( w_{\tau }\right) d\tau
\end{equation*}
where $m\left( x\right) :=1/\left( g^{2}\varphi ^{\prime }\right) \left(
x\right) $ is the (positive) speed density at $x=\varphi ^{-1}\left(
y\right) $. Both the scale function $\varphi $ and the speed measure $d\mu
=m\left( x\right) \cdot dx$ are therefore essential ingredients to reduce
the original stochastic process $\left( x_{t};t\geq 0\right) $ to the
standard Brownian motion $\left( w_{\theta };\theta \geq 0\right) $. The
Kolmogorov backward infinitesimal generator $G$ may then be written in
Feller form 
\begin{equation*}
G\left( \cdot \right) =\frac{1}{2}\frac{d}{d\mu }\left( \frac{d}{d\varphi }%
\cdot \right) .
\end{equation*}
\newline

\textbf{Examples (}from population genetics):

$\bullet $ Assume $f\left( x\right) =0$ and $g^{2}\left( x\right) =x\left(
1-x\right) $. This is the neutral WF model discussed at length later. This
diffusion is already in natural scale and $\varphi \left( x\right) =x$, $%
m\left( x\right) =\left[ x\left( 1-x\right) \right] ^{-1}.$ The speed
measure is not integrable.

$\bullet $ With $\pi _{1},\pi _{2}>0$, assume $f\left( x\right) =\pi
_{1}-\left( \pi _{1}+\pi _{2}\right) x$ and $g^{2}\left( x\right) =x\left(
1-x\right) $. This is the WF model with mutation. The parameters $\pi
_{1},\pi _{2}$ can be interpreted as mutation rates. The drift vanishes when 
$x=\pi _{1}/\pi $ (where $\pi :=\pi _{1}+\pi _{2}$ is the total mutation
pressure) which is an attracting point for the dynamics. Here:

$\varphi ^{\prime }\left( y\right) =\varphi ^{\prime }\left( y_{0}\right)
y^{-2\pi _{1}}\left( 1-y\right) ^{-2\pi _{2}},$ $\varphi \left( x\right)
=\varphi \left( x_{0}\right) +\varphi ^{\prime }\left( y_{0}\right)
\int_{x_{0}}^{x}y^{-2\pi _{1}}\left( 1-y\right) ^{-2\pi _{2}}dy,$ with $%
\varphi \left( 0\right) =-\infty $ and $\varphi \left( 1\right) =+\infty $
if $\pi _{1},\pi _{2}>1/2.$ The speed measure density is $m\left( x\right)
\propto x^{2\pi _{1}-1}\left( 1-x\right) ^{2\pi _{2}-1}$ and so is always
integrable. After normalization to $1$, $m\left( x\right) $ is the beta$%
\left( 2\pi _{1},2\pi _{2}\right) $ density.

$\bullet $ With $\sigma \in \mathbf{R},$ assume a model with quadratic
logistic drift $f\left( x\right) =\sigma x\left( 1-x\right) $ and local
variance $g^{2}\left( x\right) =x\left( 1-x\right) $. This is the WF model
with selection or selection. For this diffusion (see \cite{Kim3}), $\varphi
\left( x\right) =\frac{1-e^{-2\sigma x}}{1-e^{-2\sigma }}$ and $m\left(
x\right) \propto \left[ x\left( 1-x\right) \right] ^{-1}e^{2\sigma x}$ is
not integrable$.$ Here, $\sigma $ is a selection or fitness parameter. 
\newline

\textbf{Time change and subordination}. We start from the diffusion $\left(
1\right) $ with infinitesimal generator $G=f\partial _{x}+\frac{1}{2}%
g^{2}\partial _{x}^{2}$ and consider the time change problem without passing
first in natural coordinate$.$ Let the random time change 
\begin{equation*}
t\rightarrow \theta _{t}=\int_{0}^{t}g^{2}\left( x_{s}\right) ds.
\end{equation*}
Its inverse: $\theta \rightarrow t_{\theta }$ defined by $\theta _{t_{\theta
}}=\theta $ is given by $\theta =\int_{0}^{t_{\theta }}g^{2}\left(
x_{s}\right) ds.$

In this new stochastic time clock, the subordinated diffusion $\left(
y_{\theta }:=x_{t_{\theta }};\theta \geq 0\right) $ obeys the Langevin SDE
with potential $U\left( y\right) :=-2\int_{0}^{y}\frac{f\left( z\right) }{%
g^{2}\left( z\right) }dz$%
\begin{equation*}
dy_{\theta }=\frac{f}{g^{2}}\left( y_{\theta }\right) d\theta +dw_{\theta },
\end{equation*}
with backward infinitesimal generator $\widetilde{G}=g^{-2}G=\frac{f}{g^{2}}%
\partial _{x}+\frac{1}{2}\partial _{x}^{2}$ (See \cite{IMK}, pp. $164$-$169$%
).

We have $\overset{\cdot }{\theta }_{t}=g^{2}\left( x_{t}\right) $ meaning
that at each point $x_{t}$ of the former motion, the motion of the path is
accelerated or decelerated, depending on the rate $g^{2}\left( x_{t}\right)
\lessgtr 1.$ Note that conversely $\overset{\cdot }{t}_{\theta
}=1/g^{2}\left( y_{\theta }\right) .$ Under the time substitutions, the road
maps of the paths of both $\left( x_{t};t\geq 0\right) $ and $\left(
y_{\theta };\theta \geq 0\right) $ remain exactly the same. If a path of the
former process is accelerated or decelerated by its squared volatility $%
g^{2} $ (its local variance) at each locality$,$ then this process boils
down to the latter one. Stated differently, if we measure time by the amount
of squared volatility accumulated within each of its path, the process $%
\left( x_{t};t\geq 0\right) $ becomes $\left( y_{\theta };\theta \geq
0\right) $, both with state-space $I.$

\subsection{The transition probability density}

Assume that $f\left( x\right) $ and $g\left( x\right) $ are now
differentiable in $I$. Let then $p\left( x;t,y\right) $ stand for the
transition probability density function of $x_{t}$ at $y$ given $x_{0}=x$.
Then $p:=p\left( x;t,y\right) $ is the smallest solution to the Kolmogorov
forward (Fokker-Planck) equation (KFE): 
\begin{equation}
\Bbb{\partial }_{t}p=G^{*}\left( p\right) \text{, }p\left( x;0,y\right)
=\delta _{y}\left( x\right)  \label{2}
\end{equation}
where $G^{*}\left( \cdot \right) =-\partial _{y}\left( f\left( y\right)
\cdot \right) +\frac{1}{2}\partial _{y}^{2}\left( g^{2}\left( y\right) \cdot
\right) $ is the adjoint of $G$ ($G^{*}$ acts on the terminal value $y$
whereas $G$ acts on the initial value $x$). The way one can view this
partial differential equation (PDE) depends on the type of boundaries that $%
\left\{ 0,1\right\} $ are.

Suppose for example that the boundaries $\circ :=0$ or $1$ are both exit (or
absorbing) boundaries. From the Feller classification of boundaries, this
will be the case if $\forall y_{0}\in \left( 0,1\right) $: 
\begin{equation}
\left( i\right) \text{ }m\left( y\right) \notin L_{1}\left( y_{0},\circ
\right) \text{ and }\left( ii\right) \text{ }\varphi ^{\prime }\left(
y\right) \int_{y_{0}}^{y}m\left( z\right) dz\in L_{1}\left( y_{0},\circ
\right) ,  \label{3}
\end{equation}
where a function $f\left( y\right) \in L_{1}\left( y_{0},\circ \right) $ if $%
\int_{y_{0}}^{\circ }\left| f\left( y\right) \right| dy<+\infty $.

In this case, a sample path of $\left( x_{t};t\geq 0\right) $ can reach $%
\circ $ from the inside of $I$ in finite time but cannot reenter. The sample
paths are absorbed at $\circ $. There is an absorption at $\circ $ at time $%
\tau _{x,\circ }=\inf \left( t>0:x_{t}=\circ \mid x_{0}=x\right) $ and $%
\mathbf{P}\left( \tau _{x,\circ }<\infty \right) =1.$ Whenever both
boundaries $\left\{ 0,1\right\} $ are absorbing, the diffusion $x_{t}$
should be stopped at $\tau _{x}:=\tau _{x,0}\wedge \tau _{x,1}.$ Would none
of the boundaries $\left\{ 0,1\right\} $ be absorbing, then $\tau
_{x}=+\infty .$ This occurs when the boundaries are inaccessible.

Examples of diffusion with exit boundaries are the WF model and the WF model
with selection. In the WF model including mutations, the boundaries are
entrance boundaries and so are not absorbing.

When the boundaries are absorbing, then $p\left( x;t,y\right) $ is a
sub-probability. Letting $\rho _{t}\left( x\right) :=\int_{0}^{1}p\left(
x;t,y\right) dy$, we clearly have $\rho _{t}\left( x\right) =\mathbf{P}%
\left( \tau _{x}>t\right) $. Such models are non-conservative.

For one-dimensional diffusions, the transition density $p\left( x;t,y\right) 
$ is reversible with respect to the speed density (\cite{Kar}, Chapter $15$,
Section $13$) and so detailed balance holds: 
\begin{equation}
m\left( x\right) p\left( x;t,y\right) =m\left( y\right) p\left( y;t,x\right) 
\text{, }0<x,y<1.  \label{4}
\end{equation}
The speed density $m\left( y\right) $ satisfies $G^{*}\left( m\right) =0.$
It may be written as a Gibbs measure with density: $m\left( y\right) \propto 
\frac{1}{g^{2}\left( y\right) }e^{-U\left( y\right) }$ where the potential
function $U\left( y\right) $ reads: 
\begin{equation}
U\left( y\right) =-2\int_{0}^{y}\frac{f\left( z\right) }{g^{2}\left(
z\right) }dz\text{, }0<y<1  \label{5}
\end{equation}
and with the measure $\frac{dy}{g^{2}\left( y\right) }$ standing for the
reference measure.

Furthermore, if $p\left( s,x;t,y\right) $ is the transition probability
density from $\left( s,x\right) $ to $\left( t,y\right) $, $s<t$, then $-%
\Bbb{\partial }_{s}p=G\left( p\right) $, with terminal condition $p\left(
t,x;t,y\right) =\delta _{y}\left( x\right) $ and so $p\left( s,x;t,y\right) $
also satisfies the KBE when looking at it backward in time. The Feller
evolution semigroup being time-homogeneous, one may as well observe that
with $p:=p\left( x;t,y\right) $, operating the time substitution $%
t-s\rightarrow t$, $p$ itself solves the KBE 
\begin{equation}
\Bbb{\partial }_{t}p=G\left( p\right) ,\text{ }p\left( x;0,y\right) =\delta
_{y}\left( x\right) .  \label{6}
\end{equation}
In particular, integrating over $y$, $\Bbb{\partial }_{t}\rho _{t}\left(
x\right) =G\left( \rho _{t}\left( x\right) \right) $, with $\rho _{0}\left(
x\right) =\mathbf{1}\left( x\in \left( 0,1\right) \right) $.

$p\left( x;t,y\right) $ being a sub-probability, we may define the
normalized conditional probability density $q\left( x;t,y\right) :=p\left(
x;t,y\right) /\rho _{t}\left( x\right) $, now with total mass $1$. We get 
\begin{equation*}
\Bbb{\partial }_{t}q=-\Bbb{\partial }_{t}\rho _{t}\left( x\right) /\rho
_{t}\left( x\right) \cdot q+G^{*}\left( q\right) \text{, }q\left(
x;0,y\right) =\delta _{y}\left( x\right) .
\end{equation*}
The term $b_{t}\left( x\right) :=-\Bbb{\partial }_{t}\rho _{t}\left(
x\right) /\rho _{t}\left( x\right) >0$ is the time-dependent birth rate at
which mass should be created to compensate the loss of mass of the original
process due to absorption of $\left( x_{t};t\geq 0\right) $ at the
boundaries. In this creation of mass process, a diffusing particle started
in $x$ dies at rate $b_{t}\left( x\right) $ at point $\left( t,y\right) $
where it is duplicated in two new independent particles both started at $y$
(resulting in a global birth) evolving in the same diffusive way \footnote{%
Consider a diffusion process with forward infinitesimal generator $G^{*}$
governing the evolution of $p\left( x;t,y\right) .$ Suppose that a sample
path of this process has some probability that it will be killed or create a
new copy of itself, and that the killing and birth rates $d$ and $b$ depend
on the current location $y$ of the path. Then the process with the birth and
death opportunities of a path has the infinitesimal generator $\lambda
\left( y\right) \cdot +G^{*}\left( \cdot \right) ,$ where $\lambda \left(
y\right) =b\left( y\right) -d\left( y\right) $. The rate can also depend on $%
t$ and $x$.}. The birth rate function $b_{t}\left( x\right) $ depends here
on $x$ and $t$, not on $y.$

When the boundaries of $x_{t}$ are absorbing, the spectra of both $-G$ and $%
-G^{*}$ are discrete (see \cite{Kar} pp. $330$): There exist positive
eigenvalues $\left( \lambda _{k}\right) _{k\geq 1}$ ordered in ascending
sizes and eigenvectors $\left( v_{k},u_{k}\right) _{k\geq 1}$ of both $%
-G^{*} $ and $-G$ satisfying $-G^{*}\left( v_{k}\right) =\lambda _{k}v_{k}$
and $-G\left( y_{k}\right) =\lambda _{k}u_{k}$ such that, with $\left\langle
u_{k},v_{k}\right\rangle :=\int_{0}^{1}u_{k}\left( x\right) v_{k}\left(
x\right) dx$ and $b_{k}:=\left\langle u_{k},v_{k}\right\rangle ^{-1}$, the
spectral representation 
\begin{equation}
p\left( x;t,y\right) =\sum_{k\geq 1}b_{k}e^{-\lambda _{k}t}u_{k}\left(
x\right) v_{k}\left( y\right)  \label{7}
\end{equation}
holds$.$

Let $\lambda _{1}>\lambda _{0}=0$ be the smallest non-null eigenvalue of the
infinitesimal generator $-G^{*}$ (and of $-G$)$.$ Clearly, $-\frac{1}{t}\log
\rho _{t}\left( x\right) \underset{t\rightarrow \infty }{\rightarrow }%
\lambda _{1}$ and by L' Hospital rule therefore $b_{t}\left( x\right) 
\underset{t\rightarrow \infty }{\rightarrow }\lambda _{1}$. Putting $\Bbb{%
\partial }_{t}q=0$ in the latter evolution equation, independently of the
initial condition $x$ 
\begin{equation}
q\left( x;t,y\right) \underset{t\rightarrow \infty }{\rightarrow }q_{\infty
}\left( y\right) =v_{1}\left( y\right) ,  \label{8}
\end{equation}
where $v_{1}$ is the eigenvector of $-G^{*}$ associated to $\lambda _{1}$,
satisfying $-G^{*}v_{1}=\lambda _{1}v_{1}$. The limiting probability $v_{1}/$%
norm (after a proper normalization) is called the quasi-stationary Yaglom
limit law of $\left( x_{t};t\geq 0\right) $ conditioned on being currently
alive at all time $t$ (see \cite{Y})$.$

\subsection{Additive functionals along sample paths}

Let $\left( x_{t};t\geq 0\right) $ be the diffusion model defined by (\ref{1}%
) on the interval $I$ where both endpoints are assumed absorbing (exit).
This process is thus transient and non-conservative. We wish to evaluate the
nonnegative additive quantities 
\begin{equation*}
\alpha \left( x\right) =\mathbf{E}\left( \int_{0}^{\tau _{x}}c\left(
x_{s}\right) ds+d\left( x_{\tau _{x}}\right) \right) ,
\end{equation*}
where the functions $c$ and $d$ are both assumed nonnegative on $I$ and $%
\partial I=\left\{ 0,1\right\} $. The functional $\alpha \left( x\right)
\geq 0$ solves the Dirichlet problem: 
\begin{eqnarray*}
-G\left( \alpha \right) &=&c\text{ if }x\in I \\
\alpha &=&d\text{ if }x\in \partial I,
\end{eqnarray*}
and $\alpha $ is a super-harmonic function for $G$, satisfying $-G\left(
\alpha \right) \geq 0.$\newline

\textbf{Some examples:}

\textbf{1. }Assume $c=1$ and $d=0:$ here, $\alpha =\mathbf{E}\left( \tau
_{x}\right) $ is the mean time of absorption (average time spent in $\left(
0,1\right) $ before absorption).

\textbf{2.} Whenever both $\left\{ 0,1\right\} $ are exit boundaries, it is
of interest to evaluate the probability that $x_{t}$ first hits $\left[
0,1\right] $ (say) at $1$, given $x_{0}=x$. This can be obtained by choosing 
$c=0$ and $d\left( \circ \right) =\mathbf{1}\left( \circ =1\right) .$

\textbf{3.} Let $y\in I$ and put $c=\frac{1}{2\varepsilon }\mathbf{1}\left(
x\in \left( y-\varepsilon ,y+\varepsilon \right) \right) $ and $d=0.$ As $%
\varepsilon \rightarrow 0$, $c$ converges weakly to $\delta _{y}\left(
x\right) $ and, $\alpha =:\frak{g}\left( x,y\right) =\mathbf{E}\left( \lim 
\frac{1}{2\varepsilon }\int_{0}^{\tau _{x}}\mathbf{1}_{\left( y-\varepsilon
,y+\varepsilon \right) }\left( x_{s}\right) ds\right) =\int_{0}^{\infty
}p\left( x;s,y\right) ds$ is the Green function, solution to: 
\begin{eqnarray*}
-G\left( \frak{g}\right) &=&\delta _{y}\left( x\right) \text{ if }x\in I \\
\frak{g} &=&0\text{ if }x\in \partial I.
\end{eqnarray*}
$\frak{g}$ is therefore the mathematical expectation of the local time at $y$%
, starting from $x$ (the sojourn time density at $y$). The solution is known
to be (see \cite{Kar}, pp. $198$ or \cite{Durr}, pp. $280$) 
\begin{equation*}
\frak{g}\left( x,y\right) =2m\left( y\right) \left( \varphi \left( 1\right)
-\varphi \left( y\right) \right) \frac{\varphi \left( x\right) -\varphi
\left( 0\right) }{\varphi \left( 1\right) -\varphi \left( 0\right) }\text{
if }x\leq y
\end{equation*}
\begin{equation}
\frak{g}\left( x,y\right) =2m\left( y\right) \left( \varphi \left( y\right)
-\varphi \left( 0\right) \right) \frac{\varphi \left( 1\right) -\varphi
\left( x\right) }{\varphi \left( 1\right) -\varphi \left( 0\right) }\text{
if }x\geq y.  \label{Green}
\end{equation}
The Green function is of particular interest to solve the general problem of
evaluating additive functionals $\alpha \left( x\right) $. Indeed, as is
well-known, see \cite{Kar} for example, the integral operator with respect
to the Green kernel inverts the second order operator $-G$ leading to 
\begin{eqnarray*}
\alpha \left( x\right) &=&\int_{I}\frak{g}\left( x,y\right) c\left( y\right)
dy\text{ if }x\in I \\
\alpha &=&d\text{ if }x\in \partial I.
\end{eqnarray*}
Under this form, $\alpha \left( x\right) $ appears as a potential function
and any potential function is super-harmonic. Note that for all harmonic
function $h\geq 0$ satisfying $-G\left( h\right) =0,$%
\begin{equation*}
\alpha _{h}\left( x\right) :=\int_{I}\frak{g}\left( x,y\right) c\left(
y\right) dy+h\left( x\right)
\end{equation*}
is again super-harmonic because $-G\left( \alpha _{h}\right) =c\geq 0.$

\subsection{Transformation of sample paths (Doob-transform) producing
killing and/or branching}

In the preceding Subsections, we have dealt with a given process and
recalled the various ingredients for the expectations of various quantities
of interest, summing over the history of paths. In this setup, there is no
distinction among paths with different destinations nor did we allow for
annihilation or creation of paths inside the domain before the process
reached one of the boundaries. The Doob transform of paths allows to do so.%
\newline

Consider a one-dimensional diffusion $\left( x_{t};t\geq 0\right) $ as in (%
\ref{1}). Let $p\left( x;t,y\right) $ be its transition probability. Let $%
\alpha \left( x\right) \geq 0$ as $x\in \left[ 0,1\right] .$

Define a new transformed stochastic process $(\overline{x}_{t};t\geq 0)$ by
its transition probability 
\begin{equation}
\overline{p}\left( x;t,y\right) =\frac{\alpha \left( y\right) }{\alpha
\left( x\right) }p\left( x;t,y\right) .  \label{9}
\end{equation}
In this construction of $(\overline{x}_{t};t\geq 0)$ through a change of
measure, sample paths $x\rightarrow y$ of $\left( x_{t};t\geq 0\right) $
with a large value of the ratio $\alpha \left( y\right) /\alpha \left(
x\right) $ are favored. This is a selection of paths procedure due to Doob
(see \cite{Dyn}).

The KFE for $\overline{p}$ clearly is $\Bbb{\partial }_{t}\overline{p}$ $=%
\overline{G}^{*}(\overline{p})$, with $p\left( x;0,y\right) =\delta
_{y}\left( x\right) $ and $\overline{G}^{*}(\overline{p})=\alpha \left(
y\right) G^{*}(\overline{p}/\alpha \left( y\right) ).$ The adjoint
Kolmogorov backward operator of the transformed process is therefore by
duality 
\begin{equation}
\overline{G}\left( \cdot \right) =\frac{1}{\alpha \left( x\right) }G\left(
\alpha \left( x\right) \cdot \right) .  \label{10}
\end{equation}
Developing, with $\alpha ^{\prime }\left( x\right) :=d\alpha \left( x\right)
/dx$ and $\widetilde{G}\left( \cdot \right) :=\frac{\alpha ^{\prime }}{%
\alpha }g^{2}\partial _{x}\left( \cdot \right) +G\left( \cdot \right) $, we
get 
\begin{equation}
\overline{G}\left( \cdot \right) =\frac{1}{\alpha }G\left( \alpha \right)
\cdot +\widetilde{G}\left( \cdot \right) =:\lambda \left( x\right) \cdot +%
\widetilde{G}\left( \cdot \right)  \label{11}
\end{equation}
and the new KB operator can be obtained from the latter by adding a drift
term $\frac{\alpha ^{\prime }}{\alpha }g^{2}\partial _{x}$ to the one in $G$
of the original process to form a new process $\left( \widetilde{x}%
_{t};t\geq 0\right) $ with the KB operator $\widetilde{G}$ and by killing or
branching its sample paths at rate $\lambda \left( x\right) :=G\left( \alpha
\right) /\alpha $. In others words, with $\widetilde{f}\left( x\right)
:=f\left( x\right) +\frac{\alpha ^{\prime }}{\alpha }g^{2}\left( x\right) ,$
the novel time-homogeneous SDE to consider is 
\begin{equation}
d\widetilde{x}_{t}=\widetilde{f}\left( \widetilde{x}_{t}\right) dt+g\left( 
\widetilde{x}_{t}\right) dw_{t}\text{, }\widetilde{x}_{0}=x\in \left(
0,1\right) ,  \label{12}
\end{equation}
possibly killed or branching at rate $\lambda \left( x\right) $ as soon as $%
\lambda \neq 0$. Whenever $\left( \widetilde{x}_{t};t\geq 0\right) $ is
killed, it enters conventionally into some coffin state $\left\{ \partial
\right\} $ added to the state-space.\newline

Let us look at special cases:

$\left( i\right) $ Suppose $\alpha \geq 0$ is such that $-G\left( \alpha
\right) \geq 0$ (By $\alpha \geq 0$, we mean $\alpha >0$ in $I$, possibly
with $\alpha \left( 0\right) $ or $\alpha \left( 1\right) $ equal $0$). Then 
$\alpha $ is called a super-harmonic (or excessive) function for the process
with infinitesimal generator $G.$

In this case, the rate $\lambda \left( x\right) =:-d\left( x\right) $
satisfies $\lambda \left( x\right) \leq 0$ and only killing occurs at rate $%
d\left( x\right) $. Let $\widetilde{\tau }_{x}$ be the new absorption time
at the boundaries of $\left( \widetilde{x}_{t};t\geq 0\right) $ started at $%
x $ (with $\widetilde{\tau }_{x}=\infty $ would the boundaries be
inaccessible to the new process $\widetilde{x}_{t}$). Let $\widetilde{\tau }%
_{x,\partial } $ be the killing time of $\left( \widetilde{x}_{t};t\geq
0\right) $ started at $x$ (the hitting time of $\partial $), with $%
\widetilde{\tau }_{x,\partial }=\infty $ if $G\left( \alpha \right) \equiv
0. $ Then $\overline{\tau }_{x}:=\widetilde{\tau }_{x}\wedge \widetilde{\tau 
}_{x,\partial }$ is the novel stopping time for $\left( \widetilde{x}%
_{t};t\geq 0\right) .$ The SDE for $\left( \widetilde{x}_{t};t\geq 0\right) $%
, together with its global stopping time $\overline{\tau }_{x}$ characterize
the new process $(\overline{x}_{t};t\geq 0)$ with full generator $\overline{G%
}$ to consider. \newline

$\left( ii\right) $ Suppose $\alpha \geq 0$ is such that $-G\left( \alpha
\right) \leq 0$. Then $\alpha $ is called a sub-harmonic function for the
process with generator $G.$

In this case, the rate $\lambda \left( x\right) =:b\left( x\right) $
satisfies $\lambda \left( x\right) \geq 0$ and only branching occurs at rate 
$b\left( x\right) $. The transformed process (with infinitesimal backward
generator $\overline{G}$) accounts for a branching diffusion where a
diffusing mother particle (with generator $\widetilde{G}$ and started at $x$%
) lives a random exponential time with constant rate $1.$ When the mother
particle dies, it gives birth to a spatially dependent random number $%
M\left( x\right) $ of particles$,$ with mean $\mu \left( x\right) =1+\lambda
\left( x\right) $ (where $M\left( x\right) \overset{d}{=}1+\Delta \left(
\lambda \left( x\right) \right) $ and $\Delta \left( \lambda \left( x\right)
\right) $ is a geometrically distributed random variable on $\left\{
0,1,2,...\right\} $ with mean $\lambda \left( x\right) $). Then $M\left(
x\right) $ independent daughter particles are started afresh where their
mother particle died, with the event $M\left( x\right) =0$ impossible; they
move along a diffusion governed by $\widetilde{G}$ and reproduce,
independently and so on for the subsequent particles.

$\left( iii\right) $ If $\lambda \left( x\right) =:b\left( x\right) $ is
bounded above, $\lambda $ may be put under the alternative form 
\begin{equation*}
\lambda \left( x\right) =\lambda _{*}\left( \mu \left( x\right) -1\right) ,
\end{equation*}
where $\lambda _{*}=\sup_{x\in \left[ 0,1\right] }\lambda \left( x\right) $
and $1\leq \mu \left( x\right) \leq 2.$ In this case, we can assume that $%
M\left( x\right) $ can only take the values $1$ or $2$ with probability $%
p_{1}\left( x\right) $ and $p_{2}\left( x\right) $ respectively, with $%
p_{1}\left( x\right) +p_{2}\left( x\right) =1.$ Then, $\mu \left( x\right) =%
\mathbf{E}\left( M\left( x\right) \right) =p_{1}\left( x\right)
+2p_{2}\left( x\right) =1+p_{2}\left( x\right) $ and 
\begin{equation*}
\lambda \left( x\right) =\lambda _{*}p_{2}\left( x\right) .
\end{equation*}
Note that $\mu \left( x\right) -1=p_{2}\left( x\right) >p_{1}\left( x\right)
=2-\mu \left( x\right) .$ We get a binary branching process at rate $\lambda
_{*}$ with the event to produce two particles being more likely than the one
to produce a single one, whatever is $x.$\newline

$\left( iv\right) $ Whenever $\alpha $ is such that $-G\left( \alpha \right) 
$ has no constant sign, then both killing and branching can simultaneously
occur at the death of the mother particle. $\lambda \left( x\right) $ may be
put under the form $\lambda \left( x\right) =b\left( x\right) -d\left(
x\right) $ where $b\left( x\right) $ and $d\left( x\right) $ are the birth
(branching) and death (killing) components of $\lambda \left( x\right) $.%
\newline

$\left( v\right) $ Suppose $\lambda \left( x\right) $ is bounded below and
let $\lambda _{*}=-\inf_{x\in \left[ 0,1\right] }\lambda \left( x\right) >0.$
Then one may view $\lambda $ as 
\begin{equation*}
\lambda \left( x\right) =\lambda _{*}\left( \mu \left( x\right) -1\right) ,
\end{equation*}
where $\mu \left( x\right) \geq 0.$ In this case, branching occurs at rate $%
\lambda _{*}.$ When the mother particle dies, it gives birth to a spatially
dependent random number $M\left( x\right) $ of particles (where $M\left(
x\right) \overset{d}{=}\Delta \left( \mu \left( x\right) \right) $ and $%
\Delta \left( \mu \left( x\right) \right) $ is a geometrically distributed
random variable on $\left\{ 0,1,2,...\right\} $ with mean $\mu \left(
x\right) =1+\lambda \left( x\right) /\lambda _{*}$). With $p_{m}\left(
x\right) =\mathbf{P}\left( M\left( x\right) =m\right) =p_{0}\left( x\right)
q\left( x\right) ^{m}$, $m\geq 0,$ $p_{0}\left( x\right) =\frac{1}{1+\mu
\left( x\right) }$, $q_{0}\left( x\right) =1-p_{0}\left( x\right) $%
\begin{equation*}
\lambda \left( x\right) =\lambda _{*}\left( \sum_{m\geq 1}mp_{m}\left(
x\right) -1\right) =\lambda _{*}\left( \sum_{m\geq 2}\left( m-1\right)
p_{m}\left( x\right) -p_{0}\left( x\right) \right) .
\end{equation*}
Thus, the decomposition $\lambda \left( x\right) =b\left( x\right) -d\left(
x\right) $ holds, where $b$ and $d$ can be read from 
\begin{equation*}
\lambda \left( x\right) =\lambda _{*}\left( \frac{\mu \left( x\right) ^{2}}{%
1+\mu \left( x\right) }-\frac{1}{1+\mu \left( x\right) }\right) .
\end{equation*}
\newline

$\left( vi\right) $ In some other examples, the killing/branching rate $%
\lambda =G\left( \alpha \right) /\alpha $ is bounded above and below. Then $%
\lambda $ may be put under the form 
\begin{equation*}
\lambda \left( x\right) =\lambda _{*}\left( \mu \left( x\right) -1\right) ,
\end{equation*}
where $\lambda _{*}=\sup_{x\in \left[ 0,1\right] }\left| \lambda \left(
x\right) \right| $ and $0\leq \mu \left( x\right) \leq 2.$ In this case, we
can assume that $M\left( x\right) $ can only take the values $0$ or $2$ with
probability $p_{0}\left( x\right) $ and $p_{2}\left( x\right) $
respectively, with $p_{0}\left( x\right) +p_{2}\left( x\right) =1.$ Then, $%
\mu \left( x\right) =\mathbf{E}\left( M\left( x\right) \right) =2p_{2}\left(
x\right) $ and 
\begin{equation*}
\lambda \left( x\right) =\lambda _{*}\left( 2p_{2}\left( x\right) -1\right)
=\lambda _{*}\left( p_{2}\left( x\right) -p_{0}\left( x\right) \right) ,
\end{equation*}
giving a simple decomposition of $\lambda $ in the form $\lambda \left(
x\right) =b\left( x\right) -d\left( x\right) $ with the mother particle
living a random exponential time now with constant rate $\lambda _{*}$
before giving birth to none or two descending particles (a binary branching
process)$.$ Note that $p_{0}\left( x\right) \geq $ $p_{2}\left( x\right) $
(respectively $p_{0}\left( x\right) \geq $ $p_{2}\left( x\right) $) when $%
\mu \left( x\right) \leq 1$ ($\mu \left( x\right) \geq 1$).\newline

\textbf{Examples of} $\alpha $. When $\left( x_{t};t\geq 0\right) $ is
non-conservative, consider 
\begin{equation*}
\alpha \left( x\right) =\mathbf{E}\left( \int_{0}^{\tau _{x}}c\left(
x_{s}\right) ds+d\left( x_{\tau _{x}}\right) \right) ,
\end{equation*}
where the functions $c$ and $d$ are both assumed nonnegative on $I$ and $%
\partial I=\left\{ 0,1\right\} $. Then $\alpha \geq 0$ solves the Dirichlet
equation $-G\alpha \left( x\right) =c\left( x\right) \geq 0$ on $I$ ($%
=d\left( x\right) $ on $\partial I$) and so $\alpha $ is super-harmonic or
excessive. We refer to \cite{Hui} for examples of Doob transforms based on
such super-harmonic functions allowing to understand various conditionings
of interest when the starting point process $\left( x_{t};t\geq 0\right) $
is a neutral WF diffusion or a WF diffusion with selection.

Whenever $\alpha $ is super-harmonic for $G,$ then $\beta =1/\alpha \geq 0$
is sub-harmonic for $\widetilde{G}=G+\frac{\alpha ^{\prime }}{\alpha }%
g^{2}\partial _{x}.$ This results from the obvious identity 
\begin{equation*}
\beta ^{-1}\widetilde{G}\left( \beta \right) =-\alpha ^{-1}G\left( \alpha
\right) ,
\end{equation*}
showing that $-G\alpha \geq 0$ entails $-\widetilde{G}\left( \beta \right)
\leq 0$.\newline

Whenever $\left( x_{t};t\geq 0\right) $ is conservative and ergodic 
\begin{equation*}
\frac{1}{t}\mathbf{E}_{x}\int_{0}^{t}c\left( x_{s}\right) ds\underset{%
t\rightarrow \infty }{\rightarrow }\mu \left( c\right) :=\int_{0}^{1}c\left(
y\right) d\mu \left( y\right)
\end{equation*}
where $d\mu =m\left( y\right) dy$ is the invariant probability measure of $%
\left( x_{t};t\geq 0\right) .$ Define 
\begin{equation*}
\underset{t\rightarrow \infty }{\lim }\mathbf{E}_{x}\int_{0}^{t}c\left(
x_{s}\right) ds-t\mu \left( c\right) =\alpha \left( x\right) .
\end{equation*}
Thus 
\begin{equation*}
\alpha \left( x\right) :=\int_{0}^{\infty }\left( \mathbf{E}_{x}\left(
c\left( x_{s}\right) \right) -\mu \left( c\right) \right) ds
\end{equation*}
solves the Poisson equation 
\begin{equation*}
-G\alpha \left( x\right) =\widetilde{c}\left( x\right) :=c\left( x\right)
-\mu \left( c\right) .
\end{equation*}
We conclude that $\alpha $ is $G-$super-harmonic if ever $c\left( x\right)
\geq \mu \left( c\right) $, $\forall x$. $\diamond $\newline

\textbf{Background (multiplicative functional and path integral).} The Doob
transforms used here are particular instances of more general
transformations based on multiplicative functionals. Let $x_{t}$ be the
diffusion process (\ref{1}) governed by $G=f\partial _{x}+\frac{1}{2}%
g^{2}\partial _{x}^{2}$ with $x_{0}=x$.

Define the multiplicative functional $M_{t}$ as the solution of the
differential equation 
\begin{equation*}
dM_{t}=M_{t}\cdot \left( a\left( x_{t}\right) dt+b\left( x_{t}\right)
dw_{t}\right) ,\text{ }M_{0}=1,
\end{equation*}
where $a$ and $b$ are arbitrary twice differentiable functions. Integrating,
we get 
\begin{equation*}
M_{t}=e^{\int_{0}^{t}\left( a-\frac{1}{2}b^{2}\right) \left( x_{s}\right)
ds+\int_{0}^{t}b\left( x_{s}\right) dw_{s}}.
\end{equation*}
Let $B$ be a Borel subset of $I$. Define a new process whose density $%
\overline{p}$ is obtained after a modification of the original one while
using the multiplicative modulation factor $M_{t}$ as 
\begin{equation*}
\int_{B}\overline{p}\left( x;t,y\right) dy:=\mathbf{E}_{x}\left[ M_{t}%
\mathbf{1}\left( x_{t}\in B\right) \right] =\int_{B}\mathbf{E}_{x}\left[
M_{t}\mid x_{t}=y\right] p\left( x;t,y\right) dy.
\end{equation*}
Integrating $M_{t}$ over paths with fixed two endpoints $x$ and $y$, $%
\mathbf{E}_{x}\left[ M_{t}\mid x_{t}=y\right] $ can be interpreted as the
Radon-Nykodym derivative of $\overline{p}$ with respect to $p$, the density
of $x_{t}$. By duality, let 
\begin{equation*}
v\left( x,t\right) =\mathbf{E}_{x}\left[ M_{t}\psi \left( x_{t}\right)
\right] \text{, }v\left( x,0\right) =\psi \left( x\right) .
\end{equation*}
Applying It\^{o} calculus, we get 
\begin{equation*}
\partial _{t}v=\overline{G}\left( v\right) =\left( G+gb\partial
_{x}+a\right) \left( v\right) =:\left( \widetilde{G}+a\right) \left(
v\right) ,
\end{equation*}
where the modified backward infinitesimal generator $\overline{G}$ is
obtained by adding a drift term $gb\partial _{x}$ to $G$ to produce $%
\widetilde{G}$ and a multiplicative part $a\cdot $. The adjoint KFE giving
the evolution of $\overline{p}$ is thus 
\begin{equation*}
\partial _{t}\overline{p}=\overline{G}^{*}\left( \overline{p}\right) =\left( 
\widetilde{G}^{*}+a\right) \left( \overline{p}\right) \text{, }\overline{p}%
\left( x;0,y\right) =\delta _{y}\left( x\right) .
\end{equation*}
\newline

- (Cameron-Martin-Girsanov) For instance, when $a=0$ and $b=-f/g$, the
generator of the transformed diffusion is $\overline{G}=\frac{1}{2}%
g^{2}\partial _{x}^{2}$ killing the drift term of the original process
governed by $G.$ In this case, 
\begin{equation*}
M_{t}=e^{-\frac{1}{2}\int_{0}^{t}\left[ \left( \frac{f}{g}\right) ^{2}\left(
x_{s}\right) ds+2\frac{f}{g}\left( x_{s}\right) dw_{s}\right] }.
\end{equation*}
Clearly in this case $M_{t}$ is a martingale with $\mathbf{E}_{x}\left(
M_{t}\right) =1$, assuming $b$ to be bounded. This construction kills the
drift of the original process while using a change of measure. \newline

- (Feynman-Kac) When $b=0$, the generator of the transformed diffusion is $%
\overline{G}=G+a$ adding a multiplicative component $a$ to the one $G$
governing the original process$.$ In this case 
\begin{equation*}
M_{t}=e^{\int_{0}^{t}a\left( x_{s}\right) ds}
\end{equation*}
is the exponential of the integrated rate. If $v\left( x,t\right) =\mathbf{E}%
_{x}\left[ M_{t}\psi \left( x_{t}\right) \right] $, $v\left( x,0\right)
=\psi \left( x\right) ,$ then $v$ solves 
\begin{equation*}
\partial _{t}v=\overline{G}\left( v\right) =\left( G+a\right) \left(
v\right) ,\text{ }v\left( x,0\right) =\psi \left( x\right) .
\end{equation*}
In particular, if $v\left( x,t\right) =\mathbf{E}_{x}\left[ M_{t}\right] $, $%
v\left( x,0\right) =\mathbf{1}\left( x\in \left( 0,1\right) \right) ,$ then $%
v$ solves 
\begin{equation*}
\partial _{t}v=\overline{G}\left( v\right) =\left( G+a\right) \left(
v\right) ,\text{ }v\left( x,0\right) =\mathbf{1}\left( x\in \left(
0,1\right) \right) .
\end{equation*}
\newline

- (Doob) Suppose now 
\begin{equation*}
dM_{t}=M_{t}\left( \alpha ^{-1}\left( x_{t}\right) d\alpha \left(
x_{t}\right) \right) ,\text{ }M_{0}=1.
\end{equation*}
This $M_{t}$ is a particular instance of the general $M_{t}$ introduced
above. Indeed, applying It\^{o} calculus, 
\begin{equation*}
\alpha ^{-1}\left( x_{t}\right) d\alpha \left( x_{t}\right) =\alpha
^{-1}\alpha ^{\prime }\left[ fdt+gdw\right] +\frac{1}{2}\alpha ^{-1}\alpha
^{\prime \prime }g^{2}dt,
\end{equation*}
leading to 
\begin{eqnarray*}
a &=&\alpha ^{-1}\left( f\alpha ^{\prime }+\frac{g^{2}}{2}\alpha ^{\prime
\prime }\right) =G\left( \alpha \right) /\alpha =:\lambda \left( x\right) \\
b &=&\alpha ^{-1}\alpha ^{\prime }g.
\end{eqnarray*}
Thus $\overline{G}=G+gb\partial _{x}+a=G+\alpha ^{-1}\alpha ^{\prime
}g^{2}\partial _{x}+\lambda \left( x\right) $ as already observed earlier.

Now, from the differential generation of $M_{t}$, 
\begin{equation*}
M_{t}=\frac{\alpha \left( x_{t}\right) }{\alpha \left( x\right) },\text{ }%
M_{0}=1
\end{equation*}
only depends on the terminal and initial values of $\left( x_{s};0\leq s\leq
t\right) $ and not on its intermediate values (such a particular Doob
transformation is thus a gauge). Thus here $\mathbf{E}_{x}\left[ M_{t}\mid
x_{t}=y\right] =\frac{\alpha \left( y\right) }{\alpha \left( x\right) }$
consistently with the definition $\overline{p}\left( x;t,y\right) =\frac{%
\alpha \left( y\right) }{\alpha \left( x\right) }p\left( x;t,y\right) .$ 
\newline

\textbf{A super-harmonic example.}

Although this work chiefly focuses on Doob-transforms where branching is
present in $\lambda ,$ let us give a significant example where the Doob
transform just produces killing like in $\left( i\right) $. Suppose $\left(
x_{t};t\geq 0\right) $ is a non-conservative diffusion. Let $\lambda _{1}$
be the smallest non-null eigenvalue of the infinitesimal generator $G$ of $%
\left( x_{t};t\geq 0\right) .$ Let $\alpha =u_{1}$ be the corresponding
eigenvector, that is satisfying $-Gu_{1}=\lambda _{1}u_{1}\geq 0$ with
boundary conditions $u_{1}\left( 0\right) =u_{1}\left( 1\right) =0.$ Then $%
c=\lambda _{1}u_{1}.$ The new KB operator associated to the transformed
process $(\overline{x}_{t};t\geq 0)$ is 
\begin{equation}
\overline{G}\left( \cdot \right) =\frac{1}{\alpha }G\left( \alpha \right)
\cdot +\widetilde{G}\left( \cdot \right) =-\lambda _{1}\cdot +\widetilde{G}%
\left( \cdot \right) ,  \label{21}
\end{equation}
obtained while killing the sample paths of the process $\left( \widetilde{x}%
_{t};t\geq 0\right) $ governed by $\widetilde{G}$ at constant death rate $%
d=\lambda _{1}$. The transition probability of the transformed stochastic
process $(\overline{x}_{t};t\geq 0)$ is 
\begin{equation*}
\overline{p}\left( x;t,y\right) =\frac{u_{1}\left( y\right) }{u_{1}\left(
x\right) }p\left( x;t,y\right) .
\end{equation*}
Define $\widetilde{p}\left( x;t,y\right) =e^{\lambda _{1}t}\overline{p}%
\left( x;t,y\right) .$ It is the transition probability of the process $%
\left( \widetilde{x}_{t};t\geq 0\right) $ governed by $\widetilde{G};$ it
corresponds to the original process $\left( x_{t};t\geq 0\right) $
conditioned on never hitting the boundaries $\left\{ 0,1\right\} $ (the
so-called $Q-$process of $\left( x_{t};t\geq 0\right) $, see \cite{L}). It
is simply obtained from $\left( x_{t};t\geq 0\right) $ by adding the
additional drift term $\frac{u_{1}^{\prime }}{u_{1}}g^{2}$ to $f$, where $%
u_{1}$ is the eigenvector of $G$ associated to its smallest non-null
eigenvalue$.$ The determination of $\alpha =u_{1}$ is a Sturm-Liouville
problem. When $t$ is large, to the dominant order 
\begin{equation*}
p\left( x;t,y\right) \sim e^{-\lambda _{1}t}\frac{u_{1}\left( x\right)
v_{1}\left( y\right) }{\left\langle u_{1},v_{1}\right\rangle },
\end{equation*}
where $v_{1}$ is the Yaglom limit law of $\left( x_{t};t\geq 0\right) .$
Therefore 
\begin{equation}
\widetilde{p}\left( x;t,y\right) \sim e^{\lambda _{1}t}\frac{u_{1}\left(
y\right) }{u_{1}\left( x\right) }e^{-\lambda _{1}t}\frac{u_{1}\left(
x\right) v_{1}\left( y\right) }{\left\langle u_{1},v_{1}\right\rangle }=%
\frac{u_{1}\left( y\right) v_{1}\left( y\right) }{\left\langle
u_{1},v_{1}\right\rangle }.  \label{22}
\end{equation}
Thus the limit law of the $Q-$process $\left( \widetilde{x}_{t};t\geq
0\right) $ is the normalized Hadamard product of the eigenvectors $u_{1}$
and $v_{1}$ associated respectively to $G$ and $G^{*}.$ On the other hand,
the limit law of $\left( \widetilde{x}_{t};t\geq 0\right) $ is directly
given by 
\begin{equation}
\widetilde{p}\left( x;t,y\right) \underset{t\rightarrow \infty }{\rightarrow 
}\widetilde{p}\left( y\right) =\frac{1}{Zg^{2}\left( y\right) }%
e^{2\int_{0}^{y}\frac{f\left( z\right) +\left( \frac{u_{1}^{\prime }}{u_{1}}%
g^{2}\right) \left( z\right) }{g^{2}\left( z\right) }dz}=\frac{%
u_{1}^{2}\left( y\right) }{Zg^{2}\left( y\right) }e^{2\int_{0}^{y}\frac{%
f\left( z\right) }{g^{2}\left( z\right) }dz},  \label{23}
\end{equation}
where $Z$ is the appropriate normalizing constant. Comparing (\ref{22}) and (%
\ref{23}) 
\begin{equation*}
v_{1}\left( y\right) =\frac{u_{1}\left( y\right) }{g^{2}\left( y\right) }%
e^{2\int_{0}^{y}\frac{f\left( z\right) }{g^{2}\left( z\right) }%
dz}=u_{1}\left( y\right) m\left( y\right) .
\end{equation*}
The eigenvector $v_{1}$ associated to $G^{*}$ is therefore equal to the
eigenvector $u_{1}$ associated to $G$ times the speed density of $\left(
x_{t};t\geq 0\right) .$\newline

When dealing for example with the neutral WF diffusion (see Section $4$ for
additional details), it is known that $\lambda _{1}=1$ with $u_{1}=x\left(
1-x\right) $ and $v_{1}\equiv 1.$ The $Q-$process $\left( \widetilde{x}%
_{t};t\geq 0\right) $ in this case obeys 
\begin{equation}
d\widetilde{x}_{t}=\left( 1-2\widetilde{x}_{t}\right) dt+\sqrt{\widetilde{x}%
_{t}\left( 1-\widetilde{x}_{t}\right) }dw_{t},  \label{23a}
\end{equation}
with an additional stabilizing drift toward $1/2$: $\widetilde{f}\left(
x\right) =\frac{u_{1}^{\prime }}{u_{1}}g^{2}\left( x\right) =1-2x.$

The limit law of the $Q-$process $\left( \widetilde{x}_{t};t\geq 0\right) $
in this case is $6y\left( 1-y\right) $. The latter conditioning is more
stringent than the Yaglom conditioning and so the limiting law has more mass
away from the boundaries (compare with the uniform quasi-stationary Yaglom
limit (\ref{8}) with $v_{1}\equiv 1$).

\section{The Wright-Fisher example}

In this Section, we briefly and informally recall that the celebrated WF
diffusion process with or without a drift may be viewed as a scaling limit
of a simple two alleles discrete space-time branching process preserving the
total number $N$ of individuals in the subsequent generations (see \cite{Kar}%
, \cite{EK}, for example).

\subsection{The neutral Wright-Fisher model}

Consider a discrete-time Galton Watson branching process preserving the
total number of individuals in each generation. We start with $N$
individuals. The initial reproduction law is defined as follows: Let $\left| 
\mathbf{k}_{N}\right| :=\sum_{m=1}^{N}k_{m}=N$ and $\mathbf{k}_{N}:=\left(
k_{1},...,k_{N}\right) $ be integers. Assume the first-generation random
offspring numbers $\mathbf{\nu }_{N}:=\left( \nu _{N}\left( 1\right)
,...,\nu _{N}\left( N\right) \right) $ admit the following joint
exchangeable polynomial distribution on the discrete simplex $\left| \mathbf{%
k}_{N}\right| =N$: 
\begin{equation*}
\mathbf{P}\left( \mathbf{\nu }_{N}=\mathbf{k}_{N}\right) =\frac{N!\cdot
N^{-N}}{\prod_{n=1}^{N}k_{n}!}.
\end{equation*}
This distribution can be obtained by conditioning $N$ independent Poisson
distributed random variables on summing to $N$. Assume subsequent iterations
of this reproduction law are independent so that the population is with
constant size for all generations.

Let $N_{r}\left( n\right) $ be the offspring number of the $n$ first
individuals at the discrete generation $r\in \mathbf{N}_{0}$ corresponding
to (say) allele $A_{1}$ (the remaining number $N-N_{r}\left( n\right) $
counts the number of alleles $A_{2}$ at generation $r$). This sibship
process is a discrete-time Markov chain with binomial transition probability
given by: 
\begin{equation*}
\mathbf{P}\left( N_{r+1}\left( n\right) =k^{\prime }\mid N_{r}\left(
n\right) =k\right) =\binom{N}{k^{\prime }}\left( \frac{k}{N}\right)
^{k^{\prime }}\left( 1-\frac{k}{N}\right) ^{N-k^{\prime }}.
\end{equation*}
Assume next that $n=\left[ Nx\right] $ where $x\in \left( 0,1\right) .$
Then, as well-known, the dynamics of the continuous space-time re-scaled
process $x_{t}:=N_{\left[ Nt\right] }\left( n\right) /N$, $t\in \mathbf{R}%
_{+}$ can be approximated for large $N$, to the leading term in $N^{-1},$ by
a Wright-Fisher-It\^{o} diffusion on $\left[ 0,1\right] $ (the purely random
genetic drift case): 
\begin{equation}
dx_{t}=\sqrt{x_{t}\left( 1-x_{t}\right) }dw_{t}\text{, }x_{0}=x.  \label{28}
\end{equation}
Here $\left( w_{t};t\geq 0\right) $ is a standard Wiener process. For this
scaling limit process, a unit laps of time $t=1$ corresponds to a laps of
time $N$ for the original discrete-time process; thus time is measured in
units of $N$. If the initial condition is $x=N^{-1},$ $x_{t}$ is the
diffusion approximation of the offspring frequency of a singleton at
generation $\left[ Nt\right] $.\newline

Equation (\ref{28}) is a $1-$dimensional diffusion as in (\ref{1}) on $%
\left[ 0,1\right] ,$ with zero drift $f\left( x\right) =0$ and volatility $%
g\left( x\right) =\sqrt{x\left( 1-x\right) }$. This diffusion is already in
natural coordinate and so $\varphi \left( x\right) =x$. The scale function
is $x$ and the speed measure $\left[ x\left( 1-x\right) \right] ^{-1}dx.$
One can check that both boundaries are exit in this case: The stopping time
is $\tau _{x}=\tau _{x,0}\wedge \tau _{x,1}$ where $\tau _{x,0}$ is the
extinction time and $\tau _{x,1}$ the fixation time. The corresponding
infinitesimal generators are $G\left( \cdot \right) =\frac{1}{2}x\left(
1-x\right) \partial _{x}^{2}\left( \cdot \right) $ and $G^{*}\left( \cdot
\right) =\frac{1}{2}\partial _{y}^{2}\left( y\left( 1-y\right) \cdot \right)
.$

\subsection{Non-neutral cases}

Two alleles WF models (with non-null drifts) are classically obtained by
considering the binomial transition probabilities bin$\left( N,p_{N}\right)
: $%
\begin{equation*}
\mathbf{P}\left( N_{r+1}\left( n\right) =k^{\prime }\mid N_{r}\left(
n\right) =k\right) =\binom{N}{k^{\prime }}\left( p_{N}\left( \frac{k}{N}%
\right) \right) ^{k^{\prime }}\left( 1-p_{N}\left( \frac{k}{N}\right)
\right) ^{N-k^{\prime }}
\end{equation*}
where 
\begin{equation*}
p_{N}\left( x\right) :x\in \left( 0,1\right) \rightarrow \left( 0,1\right)
\end{equation*}
is now some state-dependent probability (which is different from the
identity $x$) reflecting some deterministic evolutionary drift from the
allele $A_{1}$ to the allele $A_{2}$. For each $r$, we have 
\begin{eqnarray*}
\mathbf{E}\left( N_{r+1}\left( n\right) \mid N_{r}\left( n\right) =k\right)
&=&Np_{N}\left( \frac{k}{N}\right) \\
\sigma ^{2}\left( N_{r+1}\left( n\right) \mid N_{r}\left( n\right) =k\right)
&=&Np_{N}\left( \frac{k}{N}\right) \left( 1-p_{N}\left( \frac{k}{N}\right)
\right)
\end{eqnarray*}
which is amenable to a diffusion approximation in terms of $x_{t}:=N_{\left[
Nt\right] }\left( n\right) /N$, $t\in \mathbf{R}_{+}$ under suitable
conditions.

For instance, taking 
\begin{equation*}
p_{N}\left( x\right) =\left( 1-\pi _{2,N}\right) x+\pi _{1,N}\left(
1-x\right) 
\end{equation*}
where $\left( \pi _{1,N},\pi _{2,N}\right) $ are small ($N$-dependent)
mutation probabilities from $A_{2}$ to $A_{1}$ (respectively $A_{1}$ to $%
A_{2}$). Assuming $\left( N\cdot \pi _{1,N},N\cdot \pi _{2,N}\right) 
\underset{}{\underset{N\rightarrow \infty }{\rightarrow }}\left( \pi
_{1},\pi _{2}\right) $, leads after scaling to the drift of WF model with
positive mutations rates $\left( \pi _{1},\pi _{2}\right) $.

Taking 
\begin{equation*}
p_{N}\left( x\right) =\frac{\left( 1+s_{1,N}\right) x}{1+s_{1,N}x+s_{2,N}%
\left( 1-x\right) }
\end{equation*}
where $s_{i,N}>0$ are small $N-$dependent selection parameter satisfying $%
N\cdot s_{i,N}\underset{N\rightarrow \infty }{\rightarrow }\sigma _{i}>0,$ $%
i=1,2,$ leads, after scaling, to the WF model with selective drift $f\left(
x\right) =\sigma x\left( 1-x\right) $, where $\sigma :=\sigma _{1}-\sigma
_{2}$. Typically, the drift $f\left( x\right) $ is a large $N$ approximation
of the bias: $N\left( p_{N}\left( x\right) -x\right) .$ The WF diffusion
with selection is thus: 
\begin{equation}
dx_{t}=\sigma x_{t}\left( 1-x_{t}\right) dt+\sqrt{x_{t}\left( 1-x_{t}\right) 
}dw_{t}  \label{29}
\end{equation}
where time is measured in units of $N.$ Letting $\theta _{t}=Nt$ define a
new time-scale with inverse $t_{\theta }=\theta /N$, the time-changed
process $y_{\theta }=x_{\theta /N}$ now obeys the SDE 
\begin{equation*}
dy_{\theta }=sy_{\theta }\left( 1-y_{\theta }\right) d\theta +\sqrt{\frac{1}{%
N}y_{\theta }\left( 1-y_{\theta }\right) }dw_{\theta },
\end{equation*}
with a small diffusion term. Here $s=s_{1}-s_{2}$ and time $\theta $ is the
usual time-clock.

The WF diffusion with selection (\ref{29}) tends to drift to $\circ =1$
(respectively $\circ =0$) if allele $A_{1}$ is selectively advantageous over 
$A_{2}:$ $\sigma _{1}>\sigma _{2}$ (respectively $\sigma _{1}<\sigma _{2}$)
in the following sense: if $\sigma >0$ (respectively $<0$), the fixation
probability at $\circ =1$, which is \cite{Kim3} 
\begin{equation*}
\mathbf{P}\left( \tau _{x,1}<\tau _{x,0}\right) =\frac{1-e^{-2\sigma x}}{%
1-e^{-2\sigma }},
\end{equation*}
increases (decreases) with $\sigma $ taking larger (smaller) values$.$

The usual way to look at the WF diffusion with mutation and selection is to
compose the two above mechanisms $p_{N}\left( x\right) $ corresponding to
mutation and selection respectively. In the scaling limit, one obtains the
standard WF diffusion model including mutations and selection as: 
\begin{equation}
dx_{t}=\left[ \left( \pi _{1}-\left( \pi _{1}+\pi _{2}\right) x_{t}\right)
+\sigma x_{t}\left( 1-x_{t}\right) \right] dt+\sqrt{x_{t}\left(
1-x_{t}\right) }dw_{t}.  \label{29a}
\end{equation}

\section{The neutral WF model}

In this Section, we particularize the general ideas developed in the
introductory Section $2$ to the neutral WF diffusion (\ref{28}) and draw
some straightforward conclusions most of which are known which illustrate
the use of Doob transforms.

\subsection{Explicit solutions of the neutral KFE}

As shown by Kimura in (\cite{Kim2}), the Kolmogorov forward (and backward)
equation is exactly solvable in this case, using spectral theory. The
solutions involve a series expansion in terms of eigen-functions of the KB
infinitesimal generator with discrete eigenvalues spectrum.

Let $\lambda _{k}=k\left( k+1\right) /2,$ $k\geq 0.$ There exist $%
u_{k}=u_{k}\left( x\right) $ and $v_{k}=v_{k}\left( y\right) $ solving the
eigenvalue problem: $-G\left( u_{k}\right) =\lambda _{k}u_{k}$ and $%
-G^{*}\left( v_{k}\right) =\lambda _{k}v_{k}$. With $\left\langle
v_{k},u_{k}\right\rangle =\int_{0}^{1}u_{k}\left( x\right) v_{k}\left(
x\right) dx$, the transition probability density $p\left( x;t,y\right) $ of
the neutral WF models admits the spectral expansion 
\begin{equation*}
p\left( x;t,y\right) =\sum_{k\geq 1}b_{k}e^{-\lambda _{k}t}u_{k}\left(
x\right) v_{k}\left( y\right) \text{where }b_{k}=\frac{1}{\left\langle
v_{k},u_{k}\right\rangle }
\end{equation*}
where $u_{k}\left( x\right) $ are the Gegenbauer polynomials rescaled on $%
\left[ 0,1\right] $ and normalized to have value $1$ at $x=0.$ In
particular, $u_{0}\left( x\right) =x$, $u_{1}\left( x\right) =x-x^{2}$, $%
u_{2}\left( x\right) =x-3x^{2}+2x^{3},$ $u_{3}\left( x\right)
=x-6x^{2}+10x^{3}-5x^{4},$ $u_{4}\left( x\right)
=x-10x^{2}+30x^{3}-35x^{4}+14x^{5},...$

Next, $v_{k}\left( y\right) =m\left( y\right) u_{k}\left( y\right) $ where $%
m\left( y\right) =1/\left( y\left( 1-y\right) \right) $ is the speed density
of the neutral WF diffusion. For instance, $v_{0}\left( y\right) =\frac{1}{%
1-y}$, $v_{1}\left( y\right) =1,$ $v_{2}\left( y\right) =1-2y,$ $v_{3}\left(
y\right) =1-5y+5y^{2},$ $v_{4}\left( y\right) =1-9y+21y^{2}-14y^{3},$...%
\newline

Although $\lambda _{0}=0$ really constitutes an eigenvalue, only $%
v_{0}\left( y\right) $ is not a polynomial and the spectral expansion of $p$
should start at $k=1,$ expressing that $p$ is a sub-probability. When $k\geq
1$, from their definition, the $u_{k}\left( x\right) $ polynomials satisfy $%
u_{k}\left( 0\right) =u_{k}\left( 1\right) =0$ in such a way that $%
v_{k}\left( y\right) =m\left( y\right) \cdot u_{k}\left( y\right) ,$ $k\geq
1 $ is a polynomial with degree $k-1$. \newline

The series expansion for $p\left( x;t,y\right) $ solves the KFE of the WF
model.

We have $\mathbf{P}\left( \tau _{x}>t\right) =\int_{0}^{1}\mathbf{P}\left(
x_{t}\in dy\right) $ and so 
\begin{equation*}
\rho _{t}\left( x\right) :=\mathbf{P}\left( \tau _{x}>t\right) =\sum_{k\geq
1}\frac{\int_{0}^{1}v_{k}\left( y\right) dy}{\left\langle
v_{k},u_{k}\right\rangle }e^{-\lambda _{k}t}u_{k}\left( x\right)
\end{equation*}
is the exact tail distribution of the absorption time.

Since $v_{1}\left( y\right) =1,$ to the leading order in $t,$ for large time 
\begin{equation*}
\mathbf{P}\left( x_{t}\in dy\right) =6e^{-t}\cdot x\left( 1-x\right) dy+%
\mathcal{O}\left( e^{-3t}\right)
\end{equation*}
which is independent of $y$. Integrating over $y$, $\rho _{t}\left( x\right)
:=\mathbf{P}\left( \tau _{x}>t\right) \sim 6e^{-t}\cdot x\left( 1-x\right) $
so that the conditional probability 
\begin{equation}
\mathbf{P}\left( x_{t}\in dy\mid \tau _{x}>t\right) \underset{t\rightarrow
\infty }{\sim }dy  \label{30}
\end{equation}
is asymptotically uniform in the Yaglom limit. As time passes by, given
absorption did not occur in the past, $x_{t}\overset{d}{\rightarrow }$ $%
x_{\infty }$ (as $t\rightarrow \infty $) which is a uniformly distributed
random variable on $\left[ 0,1\right] $.

\subsection{Additive functionals for the neutral WF and Doob transforms}

Let $\left( x_{t};t\geq 0\right) $ be the neutral WF diffusion model defined
by (\ref{28}) on the interval $I=\left[ 0,1\right] $ where both endpoints
are absorbing (exit). Consider the additive quantities 
\begin{equation*}
\alpha \left( x\right) =\mathbf{E}\left( \int_{0}^{\tau _{x}}c\left(
x_{s}\right) ds+d\left( x_{\tau _{x}}\right) \right) ,
\end{equation*}
where functions $c$ and $d$ are both nonnegative. With $G=\frac{1}{2}x\left(
1-x\right) \partial _{x}^{2}$, $\alpha \left( x\right) $ solves: 
\begin{eqnarray*}
-G\left( \alpha \right) &=&c\text{ if }x\in I \\
\alpha &=&d\text{ if }x\in \partial I.
\end{eqnarray*}
Therefore $\alpha $ is a super-harmonic function for $G.$

Take\textbf{\ }$c=\lim_{\varepsilon \downarrow 0}\frac{1}{2\varepsilon }%
\mathbf{1}_{\left( y-\varepsilon ,y+\varepsilon \right) }\left( x\right)
=:\delta _{y}\left( x\right) $ and $d=0$, when $y\in I:$ in this case, $%
\alpha :=\frak{g}\left( x,y\right) $ is the Green function (the mean local
time at $y$ given the process started at $x$). The solution takes the simple
form 
\begin{eqnarray*}
\frak{g}\left( x,y\right) &=&2\frac{x}{y}\text{ if }x<y \\
\frak{g}\left( x,y\right) &=&2\frac{1-x}{1-y}\text{ if }x>y.
\end{eqnarray*}
The Green function solves the above general problem of evaluating additive
functionals $\alpha \left( x\right) $: 
\begin{eqnarray*}
\alpha \left( x\right) &=&\int_{I}\frak{g}\left( x,y\right) c\left( y\right)
dy\text{ if }x\in I \\
\alpha &=&d\text{ if }x\in \partial I.
\end{eqnarray*}
There are many interesting choices of $c$ therefore leading to $\alpha ,$
allowing to compute for example the mean time till absorption for the
neutral WF diffusion, the probability to hit state\emph{\ }$1$ before $0...$
For each choice of $\alpha ,$ it is interesting to study the transformed
process $(\overline{x}_{t};t\geq 0)$ whose transition probability is given
by 
\begin{equation*}
\overline{p}\left( x;t,y\right) =\frac{\alpha \left( y\right) }{\alpha
\left( x\right) }p\left( x;t,y\right) ,
\end{equation*}
in terms of the original process transition probability $p\left(
x;t,y\right) .$ This allows for example to understand the neutral WF process
conditioned on exit at some boundary and to evaluate for this new process
interesting average additive functionals such as the mean time needed to hit
the exit boundary...For detailed similar examples arising in the context of
WF diffusions and related ones, see \cite{Hui}.

\section{The WF model with selection}

Now we briefly focus on the diffusion process (\ref{29}). Let $\left(
v_{k}\left( y\right) \right) _{k\geq 1}$ be the Gegenbauer eigen-polynomials
of the KF operator corresponding to the neutral WF diffusion (\ref{28}), so
with eigenvalues $\lambda _{k}=k\left( k+1\right) /2,$ $k\geq 1.$ Define the
oblate spheroidal wave functions on $\left[ 0,1\right] $ as 
\begin{equation}
w_{k}^{\sigma }\left( y\right) =\sum_{l\geq 1}^{\prime }f_{k}^{l}v_{l}\left(
y\right) ,  \label{33}
\end{equation}
where $f_{k}^{l}$ obey the three-term recurrence defined in \cite{Mano}. In
the latter equality, the $l$ summation is over odd (even) values if $k$ is
even (odd).

Define $v_{k}^{\sigma }\left( y\right) =e^{\sigma y}w_{k}^{\sigma }\left(
y\right) $ and $u_{k}^{\sigma }\left( x\right) =\frac{1}{m\left( x\right) }%
v_{k}^{\sigma }\left( x\right) $ where $m\left( x\right) =e^{2\sigma
x}/\left( x\left( 1-x\right) \right) $ is the speed measure density of the
WF model with selection (\ref{29})$.$

The system $\left( u_{k}^{\sigma }\left( x\right) ,v_{k}^{\sigma }\left(
x\right) \right) _{k\geq 1}$ constitute a system of eigen-functions for the
WF with selection generators $-G$ and $-G^{*}$ with eigenvalues $\lambda
_{k}^{\sigma }$ implicitly defined in \cite{Mano}, thus with $-G\left(
u_{k}^{\sigma }\right) =\lambda _{k}^{\sigma }u_{k}^{\sigma }$ and $%
-G^{*}\left( v_{k}^{\sigma }\right) =\lambda _{k}^{\sigma }v_{k}^{\sigma }.$
The eigen-function expansion of the transition probability density of the WF
model with selection is thus, \cite{Kim1}: 
\begin{equation}
p\left( x;t,y\right) =\sum_{k\geq 1}b_{k}^{\sigma }e^{-\lambda _{k}^{\sigma
}t}u_{k}^{\sigma }\left( x\right) v_{k}^{\sigma }\left( y\right)  \label{34}
\end{equation}
where $b_{k}^{\sigma }=\left\langle v_{k}^{\sigma },u_{k}^{\sigma
}\right\rangle ^{-1}.$ The WF model with selection can be viewed as a
perturbation problem of the neutral WF model (see \cite{Mar}). There exist
perturbation developments of $\lambda _{k}^{\sigma }$ around $\lambda _{k}$
with respect to $\sigma ^{2},$ \cite{Kim1}. They are valid and useful for
small $\sigma $.

The WF diffusion process $x_{t}$ with selection (\ref{29}) is
non-conservative, with finite hitting time $\tau _{x}$ of one of the
boundaries. Following the general arguments developed in Section $2$, the
Yaglom limit of $x_{t}$ conditioned on $\tau _{x}>t$ is the normalized
version of 
\begin{equation}
v_{1}^{\sigma }\left( y\right) =e^{\sigma y}w_{1}^{\sigma }\left( y\right) .
\label{35}
\end{equation}

The limit law of $x_{t}$ conditioned on never hitting the boundaries in the
remote future is the normalized version of 
\begin{equation}
u_{1}^{\sigma }\left( y\right) v_{1}^{\sigma }\left( y\right) =\frac{1}{%
m\left( y\right) }v_{1}^{\sigma }\left( x\right) ^{2}=y\left( 1-y\right)
w_{1}^{\sigma }\left( y\right) ^{2}.  \label{36}
\end{equation}
Because the latter conditioning is more stringent than the former, the
probability mass of (\ref{36}) is more concentrated inside the interval than
(\ref{35}). Compare with the statements at the end of Section $2$ concerning
the neutral WF diffusion.

\section{Doob transform of the neutral WF model: sub-critical BD}

In this Section, we define the branching WF diffusion model with selection
while applying a Doob transform to the neutral WF model, based on the
sub-harmonic additive functional $\alpha \left( x\right) =e^{\sigma x},$ say
with $\sigma >0$. We then study in detail the obtained branching process.%
\newline

The starting point is thus the neutral WF diffusion: $dx_{t}=\sqrt{%
x_{t}\left( 1-x_{t}\right) }dw_{t}$, $x_{0}=x\in \left( 0,1\right) .$

For this model, $G=\frac{1}{2}x\left( 1-x\right) \partial _{x}^{2}$ and both
boundaries are exit. With $\lambda _{k}=k\left( k+1\right) /2$, $k\geq 0,$
its transition density $p\left( x;t,y\right) $ admits the spectral
representation 
\begin{equation}
p\left( x;t,y\right) =\sum_{k\geq 1}b_{k}e^{-\lambda _{k}t}u_{k}\left(
x\right) v_{k}\left( y\right) ,  \label{39a}
\end{equation}
in terms of the Gegenbauer eigen-polynomials (see Subsection $4.1$). We
shall consider the following transformation of paths on the neutral WF
model: Let $\alpha \left( x\right) =e^{\sigma x},$ $\sigma >0$ and consider $%
\overline{G}\left( \cdot \right) =\alpha ^{-1}G\left( \alpha \cdot \right) =%
\widetilde{G}\left( \cdot \right) +b\left( x\right) \cdot .$ We now have $%
G\left( \alpha \right) =\frac{1}{2}\sigma ^{2}x\left( 1-x\right) e^{\sigma
x} $ and so $b\left( x\right) =G\left( \alpha \right) /\alpha =\frac{\sigma
^{2}}{2}x\left( 1-x\right) \geq 0.$

Note that $-G\left( \alpha \right) \leq 0$ indicating that $\alpha $ is
sub-harmonic for $G.$

In this case study, one selects sample paths of $\left( x_{t};t\geq 0\right) 
$ with large $\alpha \left( y\right) $ and we claim that this is an
alternative interesting way to introduce selection in the neutral WF
diffusion process.

The dynamics of $\left( \widetilde{x}_{t};t\geq 0\right) $ governed by $%
\widetilde{G}$ is easily seen to be the standard WF with selection dynamics (%
\ref{29}) 
\begin{equation*}
d\widetilde{x}_{t}=\sigma \widetilde{x}_{t}\left( 1-\widetilde{x}_{t}\right)
dt+\sqrt{\widetilde{x}_{t}\left( 1-\widetilde{x}_{t}\right) }dw_{t},
\end{equation*}
subject to additional quadratic branching at rate $b\left( x\right) =\frac{1%
}{2}\sigma ^{2}x\left( 1-x\right) $ inside $I$. We indeed have

\begin{equation*}
\overline{G}\left( \cdot \right) =e^{-\sigma x}G\left( e^{\sigma x}\cdot
\right) =b\left( x\right) \cdot +\widetilde{G}\left( \cdot \right) ,
\end{equation*}
where 
\begin{equation*}
\widetilde{G}=:\widetilde{f}\partial _{x}+\frac{1}{2}\widetilde{g}%
^{2}\partial _{x}^{2}=\sigma x\left( 1-x\right) \partial _{x}+\frac{1}{2}%
x\left( 1-x\right) \partial _{x}^{2}
\end{equation*}
is the KBE operator of the dynamics $\left( \widetilde{x}_{t};t\geq 0\right)
.$ Recall that $\widetilde{x}_{t}$ is transient and so hits one of the
boundaries $\left\{ 0,1\right\} $ in finite time $\widetilde{\tau }_{x}$.

To summarize, in our branching diffusion way to look at selection, we move
from the neutral WF diffusion $\left( x_{t};t\geq 0\right) $ to the standard
WF diffusion with selection $\left( \widetilde{x}_{t};t\geq 0\right) $ but
subject to additional branching at rate $b\left( x\right) .$ \newline

\textbf{Remark.} With $\beta \left( x\right) :=\alpha \left( x\right)
^{-1}=e^{-\sigma x},$ we clearly have 
\begin{equation*}
\overline{G}\left( \beta \left( x\right) \right) =0
\end{equation*}
and $\beta $ is an harmonic function for $\overline{G}$ and as a result,
Doob-transforming $\overline{G}$ by $\beta ,$ we get 
\begin{equation*}
\beta ^{-1}\overline{G}\left( \beta \cdot \right) =\left( \alpha \beta
\right) ^{-1}\overline{G}\left( \alpha \beta \cdot \right) =G\left( \cdot
\right)
\end{equation*}
which is the infinitesimal generator of the original neutral WF martingale. $%
\diamond $\newline

The birth (creating) rate $b\geq 0$ in $\overline{G}$ is bounded from above
on $\left( 0,1\right) $. It may be put into the canonical form $b\left(
x\right) =b_{*}\left( \mu \left( x\right) -1\right) $ where $b_{*}=\underset{%
x\in \left[ 0,1\right] }{\max }\left( b\left( x\right) \right) =\frac{\sigma
^{2}}{8}>0$ and 
\begin{equation}
\mu \left( x\right) =1+4x\left( 1-x\right) ,  \label{40}
\end{equation}
whose range is the interval $\left[ 1,2\right] $ as $x\in \left[ 0,1\right]
. $

The density of the transformed process is $\overline{p}\left( x;t,y\right) =%
\frac{\alpha \left( y\right) }{\alpha \left( x\right) }p\left( x;t,y\right)
. $ It is exactly known because so is $p$ is from (\ref{39a}).

The transformed process (with infinitesimal backward generator $\overline{G}$%
) accounts for a branching diffusion (BD) where a diffusing mother particle
(with generator $\widetilde{G}$ and started at $x$) lives a random
exponential time with constant rate $b_{*}.$ When the mother particle dies,
it gives birth to a spatially dependent random number $M\left( x\right) $ of
particles (with mean $\mu \left( x\right) $). $M\left( x\right) $
independent daughter particles are started where their mother particle died;
they move along a WF diffusion with selection and reproduce, independently
and so on.

Because $\mu \left( x\right) $ is bounded above by $2$ and larger than $1$
(indicating a super-critical branching process), we actually get a BD with
binary scission whose random offspring number satisfies (`w.p.' meaning
`with probability') 
\begin{equation*}
M\left( x\right) =0\text{ w.p. }p_{0}\left( x\right) =0
\end{equation*}
\begin{equation*}
M\left( x\right) =1\text{ w.p. }p_{1}\left( x\right) =2-\mu \left( x\right)
\end{equation*}
\begin{equation*}
M\left( x\right) =2\text{ w.p. }p_{2}\left( x\right) =\mu \left( x\right) -1,
\end{equation*}
with $p_{2}\left( x\right) \geq p_{1}\left( x\right) $ (the event that $2$
particles are generated in a splitting event is more probable than a single
one)$.$

For such a transformed process, the trade-off is as follows: there is a
competition between the boundaries $\left\{ 0,1\right\} $ which are
absorbing for the particle system and the number of particles $N_{t}\left(
x\right) $ in the system at each time $t$, which may grow due to binary
branching events (or remain steady when $M\left( x\right) =1$)$.$

The density $\overline{p}$ of the transformed process has the following
interpretation 
\begin{equation}
\overline{p}\left( x;t,y\right) =\mathbf{E}\left[ \sum_{n=1}^{N_{t}\left(
x\right) }p^{\left( n\right) }\left( x;t,y\right) \right] ,  \label{41}
\end{equation}
where $p^{\left( n\right) }\left( x;t,y\right) $ is the density at $\left(
t,y\right) $ of the $n$th alive particle descending from the ancestral one
(Eve), started at $x.$ In the latter formula, the sum vanishes if $%
N_{t}\left( x\right) =0.$ A particle is alive at time $t$ if it came to
birth before $t$ and has not been yet absorbed by the boundaries.

Let $\overline{\rho }_{t}\left( x\right) =\int_{\left( 0,1\right) }\overline{%
p}\left( x;t,y\right) dy$. Then $\overline{\rho }_{t}\left( x\right) $ is
the expected number of particle alive at time $t.$ We have 
\begin{equation*}
\partial _{t}\overline{\rho }_{t}\left( x\right) =\overline{G}\left( 
\overline{\rho }_{t}\left( x\right) \right) ,\text{ }\overline{\rho }%
_{0}\left( x\right) =\mathbf{1}\left( x\in \left( 0,1\right) \right) .
\end{equation*}
\textbf{Remark.} From the Feynman-Kac formula, $\overline{p}$ in (\ref{41})
is also 
\begin{equation*}
\overline{p}\left( x;t,y\right) =\mathbf{E}_{x}\left( e^{\int_{0}^{t\wedge 
\widetilde{\tau }_{x}}b\left( \widetilde{x}_{s}\right) ds}\mid \widetilde{x}%
_{t}=y\right) p\left( x;t,y\right)
\end{equation*}
and 
\begin{equation*}
\overline{\rho }_{t}\left( x\right) =\mathbf{E}_{x}\left(
e^{\int_{0}^{t\wedge \widetilde{\tau }_{x}}b\left( \widetilde{x}_{s}\right)
ds}\right) .\text{ }\diamond
\end{equation*}

But then $\overline{q}\left( x;t,y\right) :=\overline{p}\left( x;t,y\right) /%
\overline{\rho }_{t}\left( x\right) $ obeys the forward PDE 
\begin{equation*}
\partial _{t}\overline{q}\left( x;t,y\right) =\left( -\frac{\partial _{t}%
\overline{\rho }_{t}\left( x\right) }{\overline{\rho }_{t}\left( x\right) }%
+b\left( y\right) \right) \overline{q}\left( x;t,y\right) +\widetilde{G}%
^{*}\left( \overline{q}\left( x;t,y\right) \right)
\end{equation*}
as a result of $\partial _{t}\overline{p}\left( x;t,y\right) =\overline{G}%
^{*}\left( \overline{p}\left( x;t,y\right) \right) $. We have 
\begin{equation}
\overline{q}\left( x;t,y\right) =\frac{\mathbf{E}\left[
\sum_{n=1}^{N_{t}\left( x\right) }p^{\left( n\right) }\left( x;t,y\right)
\right] }{\mathbf{E}\left[ N_{t}\left( x\right) \right] }  \label{42}
\end{equation}
showing that $\overline{q}\left( x;t,y\right) $ is the average presence
density at $\left( t,y\right) $ of the system of particles all descending
from Eve started at $x.$

Clearly $-\frac{\log \overline{\rho }_{t}\left( x\right) }{t}\underset{%
t\rightarrow \infty }{\rightarrow }\lambda _{1}=1$ (and therefore also $-%
\frac{\partial _{t}\overline{\rho }_{t}\left( x\right) }{\overline{\rho }%
_{t}\left( x\right) }$ by L' Hospital rule), because 
\begin{equation*}
\overline{\rho }_{t}\left( x\right) =\frac{1}{\alpha \left( x\right) }%
\sum_{k\geq 1}b_{k}e^{-\lambda _{k}t}u_{k}\left( x\right) \int_{0}^{1}\alpha
\left( y\right) v_{k}\left( y\right) dy.
\end{equation*}
The expected number of particles in the system decays globally and
exponentially at rate $\lambda _{1}$.

The BD transformed process therefore admits an integrable Yaglom limit $%
\overline{q}_{\infty }$, solution to $-\widetilde{G}^{*}\left( \overline{q}%
_{\infty }\right) =\left( \lambda _{1}+b\left( y\right) \right) \overline{q}%
_{\infty }$ or $-\overline{G}^{*}\left( \overline{q}_{\infty }\right)
=\lambda _{1}\overline{q}_{\infty }$. With $v_{1}\left( y\right) =1,$ the
first eigenvector of $-G^{*}$ associated to the smallest positive eigenvalue 
$\lambda _{1}=1$, $\overline{q}_{\infty }$ is of the product form 
\begin{equation}
\overline{q}_{\infty }\left( y\right) =C_{*}e^{\sigma y}v_{1}\left( y\right)
=\frac{\sigma e^{\sigma y}}{e^{\sigma }-1}.  \label{43}
\end{equation}
The arbitrary multiplicative constant $C_{*}$ was chosen in such a way that $%
\overline{q}_{\infty }\left( y\right) $ is a probability.

By analogy with the Yaglom construction, this limiting probability $%
\overline{q}_{\infty }$can be called the quasi-stationary Yaglom average
density at $\left( t,y\right) $ for the BD particle system (it is also the
ground state for $\overline{G}^{*}$)$.$

There is also a natural eigenvector $\overline{\phi }_{\infty }$ of the
backward operator $-\overline{G}$, satisfying $-\overline{G}\left( \overline{%
\phi }_{\infty }\right) =\lambda _{1}\overline{\phi }_{\infty }$ (the ground
state for $\overline{G}$)$.$ It is explicitly here 
\begin{equation}
\overline{\phi }_{\infty }\left( x\right) =\frac{C}{\alpha \left( x\right) }%
u_{1}\left( x\right) =\frac{6\left( e^{\sigma }-1\right) }{\sigma }%
e^{-\sigma x}x\left( 1-x\right) .  \label{44}
\end{equation}
The arbitrary multiplicative constant $C=6/C_{*}$ was chosen in such a way
that $\int_{0}^{1}\overline{q}_{\infty }\left( y\right) \overline{\phi }%
_{\infty }\left( y\right) dy=1$. Note that the spectral structures of both $%
\overline{G}^{*}$ and $\overline{G}$ are easily obtainable from the ones of $%
G^{*}$ and $G$ thanks to the Doob transform structure.

In the terminology of \cite{Pin}, both operators $\overline{G}\left( \cdot
\right) +\lambda _{1}\cdot $ and its adjoint are critical \footnote{$%
\overline{G}\left( \cdot \right) +\lambda _{1}\cdot $ ($\overline{G}%
^{*}\left( \cdot \right) +\lambda _{1}\cdot $) is said to be critical if
there exists some function $\overline{\phi }_{\infty }\in C^{2}$
(respectively $\overline{q}_{\infty }\in C^{2}$), strictly positive in $%
\left( 0,1\right) ,$ such that: $\overline{G}\left( \overline{\phi }_{\infty
}\right) +\lambda _{1}\overline{\phi }_{\infty }=0$ (respectively $\overline{%
G}^{*}\left( \overline{q}_{\infty }\right) +\lambda _{1}\overline{q}_{\infty
}=0$) and the operators do not possess a minimal positive Green function.}.
In this context, the constant $\lambda _{1}$ is called the generalized
principal eigenvalue. The eigen-functions $\left( \overline{\phi }_{\infty },%
\overline{q}_{\infty }\right) $ are their associated ground states. We note
that we have the $L^{1}-$product property (See \cite{Pin}, Subsection $4.9$%
). 
\begin{equation*}
\int_{0}^{1}\overline{\phi }_{\infty }\left( x\right) \overline{q}_{\infty
}\left( x\right) dx=6\int_{0}^{1}u_{1}\left( x\right) v_{1}\left( x\right)
dx=1<\infty .
\end{equation*}
\newline

\textbf{Remark.} Using the Feynman-Kac representation of $\overline{\rho }%
_{t}\left( x\right) ,$ we get 
\begin{equation*}
-\frac{1}{t}\log \mathbf{E}_{x}\left( e^{\int_{0}^{t\wedge \widetilde{\tau }%
_{x}}b\left( \widetilde{x}_{s}\right) ds}\right) \underset{t\rightarrow
\infty }{\rightarrow }\lambda _{1}=1\text{ and}
\end{equation*}
\begin{equation*}
e^{\lambda _{1}t}\mathbf{E}_{x}\left( e^{\int_{0}^{t\wedge \widetilde{\tau }%
_{x}}b\left( \widetilde{x}_{s}\right) ds}\right) \underset{t\rightarrow
\infty }{\rightarrow }\overline{\phi }_{\infty }\left( x\right) .\text{ }%
\diamond
\end{equation*}
\newline

With $p_{m}\left( x\right) =\mathbf{P}\left( M\left( x\right) =m\right) $,
let 
\begin{equation*}
l\left( x\right) =\sum_{m\geq 1}p_{m}\left( x\right) m\log m=2\log
2p_{2}\left( x\right) .
\end{equation*}
We have the $x\log x$ condition: 
\begin{equation}
\int_{0}^{1}l\left( x\right) \overline{\phi }_{\infty }\left( x\right) 
\overline{q}_{\infty }\left( x\right) dx=48\log 2\int_{0}^{1}x\left(
1-x\right) u_{1}\left( x\right) v_{1}\left( x\right) dx<\infty .  \label{44a}
\end{equation}
We conclude (following \cite{AH1} and \cite{AH2}) that, as a result of the
condition (\ref{44a}) being trivially satisfied, global extinction holds in
the following sense:\newline

$\left( i\right) $ $\mathbf{P}\left( N_{t}\left( x\right) =0\right) 
\underset{t\rightarrow \infty }{\rightarrow }1$, uniformly in $x.$

$\left( ii\right) $ there exists a constant $\gamma >0:$ $e^{\lambda
_{1}t}\left[ 1-\mathbf{P}\left( N_{t}\left( x\right) =0\right) \right] 
\underset{t\rightarrow \infty }{\rightarrow }\gamma \overline{\phi }_{\infty
}\left( x\right) ,$ uniformly in $x.$

$\left( iii\right) $ For all bounded measurable function $\psi $ on $I:$ 
\begin{equation*}
\mathbf{E}\left[ \sum_{n=1}^{N_{t}\left( x\right) }\psi \left( \widetilde{x}%
_{t}^{\left( n\right) }\right) \mid N_{t}\left( x\right) >0\right] \underset{%
t\rightarrow \infty }{\rightarrow }\gamma ^{-1}\int_{\left( 0,1\right) }\psi
\left( y\right) \overline{q}_{\infty }\left( y\right) dy.
\end{equation*}
\newline

From $\left( i\right) $, it is clear that the process gets ultimately
extinct with probability $1.$ In the trade-off between pure branching and
absorption at the boundaries, all particles get eventually absorbed and the
global BD process turns out be sub-critical (even though $\mu \left(
x\right) =\mathbf{E}M\left( x\right) >1$ for all $x\in \left( 0,1\right) $):
Probability mass escapes out of $I$ although the BD survives with positive
probability.

In the statement $\left( ii\right) ,$ the quantity $1-\mathbf{P}\left(
N_{t}\left( x\right) =0\right) =\mathbf{P}\left( N_{t}\left( x\right)
>0\right) $ is also $\mathbf{P}\left( T\left( x\right) >t\right) $ where $%
T\left( x\right) $ is the global extinction time of the particle system
descending from an Eve particle started at $x$. The number $-\lambda _{1}$
is the usual Malthus exponential decay rate parameter. From $\left(
ii\right) ,$ $\overline{\phi }_{\infty }\left( x\right) $ has a natural
interpretation in terms of the propensity of the particle system to survive
to its extinction fate: the so-called reproductive value in demography.

$\left( iii\right) $ with $\psi =1$ reads $\mathbf{E}\left[ N_{t}\left(
x\right) \mid N_{t}\left( x\right) >0\right] \underset{t\rightarrow \infty }{%
\rightarrow }\gamma ^{-1}$ giving an interpretation of the constant $\gamma $
(which may be hard to evaluate in practise).\newline

The ground states of $\overline{G}+\lambda _{1}$ and its adjoint are thus $%
\left( \overline{\phi }_{\infty },\overline{q}_{\infty }\right) $ and
explicit here. It is useful to consider the process whose infinitesimal
generator is given by the Doob-transform 
\begin{equation*}
\overline{\phi }_{\infty }^{-1}\left( \overline{G}+\lambda _{1}\right)
\left( \overline{\phi }_{\infty }\cdot \right) =\overline{\phi }_{\infty
}^{-1}\left( \widetilde{G}+b+\lambda _{1}\right) \left( \overline{\phi }%
_{\infty }\cdot \right) ,
\end{equation*}
because product-criticality is preserved under this transformation. The
ground states associated to this new operator and its dual are $\left( 1,%
\overline{\phi }_{\infty }\overline{q}_{\infty }\right) $. Developing, we
obtain a process whose infinitesimal generator is 
\begin{equation*}
\widetilde{G}+\frac{\overline{\phi }_{\infty }^{\prime }}{\overline{\phi }%
_{\infty }}g^{2}\partial _{x}=G+\frac{u_{1}^{\prime }}{u_{1}}g^{2}\partial
_{x},
\end{equation*}
with no multiplicative part. In our case study, we get $\frac{1}{2}x\left(
1-x\right) \partial _{x}^{2}+\left( 1-2x\right) \partial _{x}$ adding a
stabilizing drift towards $1/2$ to the original neutral WF model$.$ The
associated diffusion process is positive recurrent and so its invariant
measure $\overline{\phi }_{\infty }\overline{q}_{\infty
}=6u_{1}v_{1}=6y\left( 1-y\right) $ is integrable with mass $1$. It is the
beta$\left( 2,2\right) $ limit law of the $Q-$process (see (\ref{23a}) and
the comments at the end of Section $2$ relative to the neutral WF diffusion).%
\newline

\textbf{Remarks.}

$\left( i\right) $ At time $t$, let $\left( \widetilde{x}_{t}^{\left(
n\right) }\right) _{n=1}^{N_{t}\left( x\right) }$ denote the positions of
the BD particle system. Let $u\left( x,t;z\right) =\mathbf{E}\left[
\prod_{n=1}^{N_{t}\left( x\right) }z^{\psi \left( \widetilde{x}_{t}^{\left(
n\right) }\right) }\right] $ stand for the functional generating function ($%
\left| z\right| \leq 1$) of the measure-valued branching particle system. $%
u\left( x,t;z\right) $ obeys the nonlinear (quadratic)
Kolmogorov-Petrovsky-Piscounoff PDE, \cite{KPP}: 
\begin{equation*}
\partial _{t}u\left( x,t;z\right) =b_{*}\theta \left( x,u\left( x,t;z\right)
\right) +\widetilde{G}\left( u\left( x,t;z\right) \right) ;\text{ }u\left(
x,0;z\right) =z^{\psi \left( x\right) },
\end{equation*}
where $\theta \left( x,z\right) =\mathbf{E}\left[ z^{M\left( x\right)
}\right] -z=\left( p_{2}\left( x\right) z^{2}+p_{1}\left( x\right) z\right)
-z$ or 
\begin{equation*}
\theta \left( x,z\right) =4x\left( 1-x\right) z\left( z-1\right)
\end{equation*}
is the shifted probability generating function of the branching law of $%
M\left( x\right) .$ Thus, the nonlinear part reads $b_{*}\theta \left(
x,u\left( x,t;z\right) \right) =b\left( x\right) u\left( x,t;z\right) \left(
u\left( x,t;z\right) -1\right) $ which is quadratic in $u.$

In particular, if $u\left( x,t\right) :=\partial _{z}u\left( x,t;z\right)
_{z=1}=\mathbf{E}\left[ \sum_{n=1}^{N_{t}\left( x\right) }\psi \left( 
\widetilde{x}_{t}^{\left( n\right) }\right) \right] $, $u\left( x,t\right) $
obeys the linear backward PDE 
\begin{equation*}
\partial _{t}u\left( x,t\right) =b\left( x\right) u\left( x,t\right) +%
\widetilde{G}\left( u\left( x,t\right) \right) ;\text{ }u\left( x,0\right)
=\psi \left( x\right)
\end{equation*}
involving $\overline{G}\left( \cdot \right) =\widetilde{G}\left( \cdot
\right) +b\left( x\right) \cdot $. We have the Feynman-Kac interpretation 
\begin{equation*}
u\left( x,t\right) =\mathbf{E}_{x}\left( e^{\int_{0}^{t\wedge \widetilde{%
\tau }_{x}}b\left( \widetilde{x}_{s}\right) ds}\psi \left( \widetilde{x}%
_{t}\right) \right) .
\end{equation*}
The latter evolution equation is the backward version of the forward PDE
giving the evolution of $\overline{p}\left( x;t,y\right) $ as $\partial _{t}%
\overline{p}\left( x;t,y\right) =\overline{G}^{*}\left( \overline{p}\left(
x;t,y\right) \right) ,$ $\overline{p}\left( x;0,y\right) =\delta _{x}\left(
y\right) .$ \newline

$\left( ii\right) $ Let us look at the branching diffusion process governed
by $\overline{G}$ would time be measured using the time substitution 
\begin{equation*}
\theta _{t}=\int_{0}^{t}g^{2}\left( \widetilde{x}_{s}\right) ds=\int_{0}^{t}%
\widetilde{x}_{s}\left( 1-\widetilde{x}_{s}\right) ds
\end{equation*}
for each of the particles that came to birth before $t$.

Then $\overline{G}\rightarrow \overline{\mathcal{G}}:=\frac{1}{x\left(
1-x\right) }\overline{G}=\sigma \partial _{x}+\frac{1}{2}\partial _{x}^{2}+%
\frac{1}{2}\sigma ^{2}\cdot .$ In particular, each motion $y_{\theta }=%
\widetilde{x}_{t_{\theta }}$ is a Brownian motion with constant drift (a
Gaussian process). This new $\overline{\mathcal{G}}$ is the one of absorbing
Brownian motion with drift $\sigma $ on $\left[ 0,1\right] ,$ including
branching at constant rate $\frac{1}{2}\sigma ^{2}.$ The Sturm-Liouville
problem for $\overline{\mathcal{G}}$ admits the eigenvalues $\lambda _{k}=%
\frac{k^{2}+\sigma ^{2}}{2}$, $k\geq 1$ with eigen-states $u_{k}\left(
x\right) \propto e^{-\sigma x}\sin \left( k\pi x\right) $ and $v_{k}\left(
y\right) \propto e^{\sigma y}\sin \left( k\pi y\right) .$ The spectral gap
is $\lambda _{1}=\frac{1+\sigma ^{2}}{2}>0$ and the time-changed branching
diffusion also becomes eventually extinct, sub-critically: The time
substitution changes the spectral structure of the model but not its
qualitative features. $\diamond $

\section{Doob transform of the WF model with mutations: critical BD}

In this Section, we start from the WF model with mutations. Using the same
Doob transform based on the additive functional $\alpha \left( x\right)
=e^{\sigma x}$ to introduce selection, we end up with a WF diffusion process
with killing and branching describing the effect of selection on the WF
model in the presence of mutations. We show that in this setup, the
resulting branching diffusion process is no longer sub-critical; rather, it
turns out to be critical.\newline

Suppose the starting point model is now the WF diffusion with mutations: 
\begin{equation*}
dx_{t}=\left( \pi _{1}-\pi x_{t}\right) dt+\sqrt{x_{t}\left( 1-x_{t}\right) }%
dw_{t},x_{0}=x\in \left( 0,1\right) ,
\end{equation*}
with $\pi :=\pi _{1}+\pi _{2}$. For this model, $G=\left( \pi _{1}-\pi
x\right) \partial _{x}+\frac{1}{2}x\left( 1-x\right) \partial _{x}^{2}$ and
both boundaries are chosen as being entrance (reflecting)\footnote{%
When both the mutation rates $u_{1}$ and $u_{2}$ are greater than $1/2$, the
boundaries are entrance. When either $u_{1}$ or $u_{2}$ is smaller than $1/2$
the corresponding boundary is regular and one needs to specify whether it is
reflecting or absorbing or a mixture of the two. We force here the regular
boundaries to be entrance.}. The WF diffusion process with mutations is now
ergodic. With 
\begin{equation*}
\lambda _{k}=\frac{k\left( k-1+\pi \right) }{2}\text{, }k\geq 0,
\end{equation*}
its transition density $p\left( x;t,y\right) $ now admits the discrete
spectral representation 
\begin{equation}
p\left( x;t,y\right) =\sum_{k\geq 0}b_{k}e^{-\lambda _{k}t}u_{k}\left(
x\right) v_{k}\left( y\right) .  \label{39aa}
\end{equation}
Here, $u_{k}\left( x\right) $ are the Jacobi polynomials rescaled on $\left[
0,1\right] $ and normalized to have value $1$ at $x=0.$ In particular, $%
u_{0}\left( x\right) =1$, $u_{1}\left( x\right) =1-\frac{\pi }{\pi _{2}}x$, $%
u_{2}\left( x\right) =1-\frac{2\left( 1+\pi \right) }{\pi _{2}}x+\frac{%
\left( 1+\pi \right) \left( 2+\pi \right) }{\pi _{2}\left( 1+\pi _{2}\right) 
}x^{2},$... Next, $v_{k}\left( y\right) =m\left( y\right) u_{k}\left(
y\right) $ where 
\begin{equation*}
m\left( y\right) =\frac{\Gamma \left( 2\pi \right) }{\Gamma \left( 2\pi
_{1}\right) \Gamma \left( 2\pi _{2}\right) }y^{2\pi _{1}-1}\left( 1-y\right)
^{2\pi _{2}-1}
\end{equation*}
is the speed density of the ergodic WF diffusion with mutations (its
normalized invariant measure). Note that the $k=0$ term in (\ref{39aa}) is
precisely $m\left( y\right) $ as required. Because the transition
probability density of the WF diffusion with mutations has also a discrete
spectral representation, this model is amenable to a similar analysis than
the neutral WF diffusion.\newline

Proceeding as for the neutral case, we shall consider the following
transformation of paths for the WF model with mutations: Let $\alpha \left(
x\right) =e^{\sigma x}$ and consider a transformed process with
infinitesimal generator $\overline{G}\left( \cdot \right) =\alpha
^{-1}G\left( \alpha \cdot \right) .$ The multiplicative part of $\overline{G}
$ is now 
\begin{equation*}
\lambda \left( x\right) =G\left( \alpha \right) /\alpha =\sigma \left( \pi
_{1}-\pi x\right) +\frac{\sigma ^{2}}{2}x\left( 1-x\right) .
\end{equation*}
Note that now $\alpha $ is neither sub-harmonic nor super-harmonic for the
infinitesimal generator $G$ including mutations because the sign of $\lambda
\left( x\right) $ varies as $x$ varies.\newline

In this case study, one selects sample paths of the WF diffusion model with
mutations $\left( x_{t};t\geq 0\right) $ with large terminal values of $%
\alpha \left( y\right) .$ The dynamics of $\left( \widetilde{x}_{t};t\geq
0\right) $ is easily seen to be the WF with mutation and selection dynamics
of the type (\ref{29a}) 
\begin{equation*}
d\widetilde{x}_{t}=\left[ \left( \pi _{1}-\pi \widetilde{x}_{t}\right)
+\sigma \widetilde{x}_{t}\left( 1-\widetilde{x}_{t}\right) \right] dt+\sqrt{%
\widetilde{x}_{t}\left( 1-\widetilde{x}_{t}\right) }dw_{t},
\end{equation*}
subject to additional quadratic killing and branching at rate $\lambda
\left( x\right) $ inside $I$. We indeed have

\begin{equation*}
\overline{G}\left( \cdot \right) =e^{-\sigma x}G\left( e^{\sigma x}\cdot
\right) =\lambda \left( x\right) \cdot +\widetilde{G}\left( \cdot \right) ,
\end{equation*}
where $\widetilde{G}=\left[ \left( \pi _{1}-\pi x\right) +\sigma x\left(
1-x\right) \right] \partial _{x}+\frac{1}{2}x\left( 1-x\right) \partial
_{x}^{2}$ is the KBE operator of the dynamics $\left( \widetilde{x}%
_{t};t\geq 0\right) .$

To summarize, in our branching diffusion way to look at the action of
selection, we move from the WF diffusion with mutations $\left( x_{t};t\geq
0\right) $ to the standard WF diffusion with mutation and selection $\left( 
\widetilde{x}_{t};t\geq 0\right) ,$ but subject to additional
killing/branching at rate $\lambda \left( x\right) .$\newline

\textbf{Remark.} With $\beta =\alpha ^{-1}=e^{-\sigma x},$ again $\overline{G%
}\left( \beta \right) =0$ and $\beta ^{-1}\overline{G}\left( \beta \cdot
\right) =G\left( \cdot \right) $ is the infinitesimal generator of the
original WF model, now with mutations. $\diamond $\newline

The birth (creating) and death (annihilating) rate $\lambda $ in $\overline{G%
}$ is bounded from above and below on $\left( 0,1\right) $. It may now be
put into the canonical form $\lambda \left( x\right) =\lambda _{*}\left( \mu
\left( x\right) -1\right) $ where $\lambda _{*}=\underset{x\in \left[
0,1\right] }{\max }\left( \left| \lambda \left( x\right) \right| \right) $%
and 
\begin{equation}
\mu \left( x\right) =1+\frac{\lambda \left( x\right) }{\lambda _{*}}
\label{39ab}
\end{equation}
whose range belongs to the interval $\left[ 0,2\right] $ as $x\in \left[
0,1\right] .$

Note that when $\pi >\sigma /2$, $\lambda _{*}=\sigma \left( \pi _{1}\vee
\pi _{2}\right) $ whereas when $\pi <\sigma /2$, $\lambda _{*}=\lambda
\left( x_{*}\right) \vee \sigma \pi _{2}$ where $x_{*}=1/2-\pi /\sigma >0.$

The density of the transformed process is $\overline{p}\left( x;t,y\right) =%
\frac{\alpha \left( y\right) }{\alpha \left( x\right) }p\left( x;t,y\right)
. $ It is exactly known because $p$ is known from (\ref{39aa}).

The transformed process (with infinitesimal backward generator $\overline{G}$%
) accounts for a branching diffusion (BD) where a diffusing mother particle
(with generator $\widetilde{G}$ and started at $x$) lives a random
exponential time with constant rate $\lambda _{*}.$ When the mother particle
dies, it gives birth to a spatially dependent random number $M\left(
x\right) $ of particles (with mean $\mu \left( x\right) $). If $M\left(
x\right) \neq 0$, $M\left( x\right) $ independent daughter particles are
started where their mother particle died; they move along a WF diffusion
with mutation and selection (with generator $\widetilde{G}$) and reproduce
independently, and so on.

Because $\mu \left( x\right) $ is bounded above by $2$ and larger than $0$,
we actually get a BD with binary scission whose random offspring number
satisfies 
\begin{equation*}
M\left( x\right) =0\text{ w.p. }p_{0}\left( x\right) =1-\mu \left( x\right)
/2
\end{equation*}
\begin{equation*}
M\left( x\right) =1\text{ w.p. }p_{1}\left( x\right) =0
\end{equation*}
\begin{equation*}
M\left( x\right) =2\text{ w.p. }p_{2}\left( x\right) =\mu \left( x\right) /2.
\end{equation*}

Note that 
\begin{equation*}
\lambda \left( x\right) =\lambda _{*}\left( p_{2}\left( x\right)
-p_{0}\left( x\right) \right) =:b\left( x\right) -d\left( x\right)
\end{equation*}
identifying the birth and death components of the full multiplicative rate $%
\lambda \left( x\right) $.

For such a transformed process, the trade-off is of a different nature:
there is a competition between the boundaries $\left\{ 0,1\right\} $ which
are now reflecting for the system of particles and the number of particles $%
N_{t}\left( x\right) $ in the system at each time $t$, which may grow or
diminish due either to branching or killing events$.$ In the presence of
mutations, the particles are no longer killed once they hit the boundaries,
suggesting that there should be a greater amount of them alive in the
system. However, in this new model, there is an opportunity to kill the
particles inside the definition domain, when they branch. The question now
being: does the new trade-off result in global extinction or global
explosion of the particle system? We will now show that critical global
extinction occurs.\newline

The density $\overline{p}$ of the transformed process again has the
interpretation (\ref{41}), where $p^{\left( n\right) }\left( x;t,y\right) $
is the density at $\left( t,y\right) $ of the $n$th alive particle
descending from the ancestral one (Eve), started at $x.$ In the latter
formula, the sum vanishes if $N_{t}\left( x\right) =0.$ A particle is alive
at time $t$ if it came to birth before $t$ and has not yet been killed by a
killing event.

Let $\overline{\rho }_{t}\left( x\right) =\int_{\left( 0,1\right) }\overline{%
p}\left( x;t,y\right) dy$. Then $\overline{\rho }_{t}\left( x\right) $ is
the expected number of particle alive at time $t.$ We have 
\begin{equation*}
\partial _{t}\overline{\rho }_{t}\left( x\right) =\overline{G}\left( 
\overline{\rho }_{t}\left( x\right) \right) ,\text{ }\overline{\rho }%
_{0}\left( x\right) =\mathbf{1}\left( x\in \left( 0,1\right) \right) .
\end{equation*}
But then $\overline{q}\left( x;t,y\right) :=\overline{p}\left( x;t,y\right) /%
\overline{\rho }_{t}\left( x\right) $ obeys the forward PDE 
\begin{equation*}
\partial _{t}\overline{q}\left( x;t,y\right) =\left( -\frac{\partial _{t}%
\overline{\rho }_{t}\left( x\right) }{\overline{\rho }_{t}\left( x\right) }%
+b\left( y\right) \right) \overline{q}\left( x;t,y\right) +\widetilde{G}%
^{*}\left( \overline{q}\left( x;t,y\right) \right)
\end{equation*}
as a result of $\partial _{t}\overline{p}\left( x;t,y\right) =\overline{G}%
^{*}\left( \overline{p}\left( x;t,y\right) \right) $. We again have (\ref{42}%
), with $\overline{q}\left( x;t,y\right) $ the average presence density at $%
\left( t,y\right) $ of the system of particles all descending from Eve
started at $x.$

Clearly $-\frac{\log \overline{\rho }_{t}\left( x\right) }{t}\underset{%
t\rightarrow \infty }{\rightarrow }\lambda _{0}=0$ (and therefore also $-%
\frac{\partial _{t}\overline{\rho }_{t}\left( x\right) }{\overline{\rho }%
_{t}\left( x\right) }$), because 
\begin{equation*}
\overline{\rho }_{t}\left( x\right) =\frac{1}{\alpha \left( x\right) }%
\sum_{k\geq 0}b_{k}e^{-\lambda _{k}t}u_{k}\left( x\right) \int_{0}^{1}\alpha
\left( y\right) v_{k}\left( y\right) dy.
\end{equation*}
The expected number of particles in the system decays globally at rate $%
\lambda _{1}$ towards the non-zero limiting value 
\begin{equation*}
\overline{\rho }_{\infty }\left( x\right) :=\alpha \left( x\right)
^{-1}b_{0}u_{0}\left( x\right) \int_{0}^{1}\alpha \left( y\right)
v_{0}\left( y\right) dy=e^{-\sigma x}\int_{0}^{1}e^{\sigma y}m\left(
y\right) dy.
\end{equation*}
The BD transformed process therefore admits an integrable Yaglom limit $%
\overline{q}_{\infty }$, solution to $-\widetilde{G}^{*}\left( \overline{q}%
_{\infty }\right) =\lambda \left( y\right) \overline{q}_{\infty }$ or $-%
\overline{G}^{*}\left( \overline{q}_{\infty }\right) =0$. With $v_{0}\left(
y\right) =m\left( y\right) ,$ the first eigenvector of $-G^{*}$ associated
to the smallest positive eigenvalue $\lambda _{0}=0$ (the equilibrium
density of the WF diffusion with mutations), $\overline{q}_{\infty }$ is of
the product form 
\begin{equation}
\overline{q}_{\infty }\left( y\right) =\frac{e^{\sigma y}m\left( y\right) }{%
\int_{0}^{1}e^{\sigma y}m\left( y\right) dy}.  \label{43a}
\end{equation}
This explicit limiting probability $\overline{q}_{\infty }$is the Yaglom
limiting average presence density at $\left( t,y\right) $ for the BD system
of particles (it is also the ground state for $\overline{G}^{*}$)$.$

There is also a natural eigenvector $\overline{\phi }_{\infty }$ of the
backward operator $-\overline{G}$, satisfying $-\overline{G}\left( \overline{%
\phi }_{\infty }\right) =0$ (the ground state for $\overline{G}$)$.$ It is
explicitly here 
\begin{equation}
\overline{\phi }_{\infty }\left( x\right) =\frac{1}{\alpha \left( x\right) }%
u_{0}\left( x\right) \int_{0}^{1}e^{\sigma y}m\left( y\right) dy=e^{-\sigma
x}\int_{0}^{1}e^{\sigma y}m\left( y\right) dy.  \label{44aa}
\end{equation}

Both operators $\overline{G}\left( \cdot \right) $ and its adjoint are again
critical. The constant $\lambda _{0}=0$ is the new generalized principal
eigenvalue; The eigen-functions $\left( \overline{\phi }_{\infty },\overline{%
q}_{\infty }\right) $ are the new associated ground states.

We note that we have the $L^{1}-$product property 
\begin{equation*}
\int_{0}^{1}u_{0}\left( x\right) v_{0}\left( x\right) dx=\int_{0}^{1}%
\overline{\phi }_{\infty }\left( x\right) \overline{q}_{\infty }\left(
x\right) dx=1<\infty .
\end{equation*}
Clearly the ground states of $-\overline{G}^{*}$ and $-\overline{G}$ are
defined up to arbitrary multiplicative constants. Note that we chose these
constants in such a way that $\int_{0}^{1}\overline{q}_{\infty }\left(
y\right) dy=1$ and $\int_{0}^{1}\overline{\phi }_{\infty }\left( x\right) 
\overline{q}_{\infty }\left( x\right) dx=1.$

With $p_{m}\left( x\right) =\mathbf{P}\left( M\left( x\right) =m\right) $,
let 
\begin{equation*}
\kappa \left( x\right) =\sum_{m\geq 2}m\left( m-1\right) p_{m}\left(
x\right) =2p_{2}\left( x\right) .
\end{equation*}
Because $p_{2}\left( x\right) $ is a degree two polynomial in $x$, we have
the condition: 
\begin{equation}
\int_{0}^{1}\kappa \left( x\right) \overline{\phi }_{\infty }\left( x\right) 
\overline{q}_{\infty }\left( x\right) dx=2\int_{0}^{1}p_{2}\left( x\right)
u_{0}\left( x\right) v_{0}\left( x\right) dx<\infty .  \label{44ab}
\end{equation}
We conclude (following \cite{AH1} and \cite{AH2}) that, as a result of the
condition (\ref{44ab}) being trivially satisfied, global extinction holds
critically, in the following sense:\newline

$\left( i\right) $ $\mathbf{P}\left( N_{t}\left( x\right) =0\right) 
\underset{t\rightarrow \infty }{\rightarrow }1$, uniformly in $x.$

$\left( ii\right) $ Let $\mu _{t}=\sum_{n=1}^{N_{t}\left( \cdot \right)
}\delta _{x_{t}^{\left( n\right) }}$, with $\mu _{t}\left( \psi \right)
=\sum_{n=1}^{N_{t}\left( \cdot \right) }\psi \left( x_{t}^{\left( n\right)
}\right) .$

There exists a finite positive constant $:$ 
\begin{equation*}
\mu =\frac{1}{2t}\int_{0}^{1}\mathbf{E}_{x}\left[ \mu _{t}\left( \phi
\right) ^{2}-\mu _{t}\left( \phi ^{2}\right) \right] \overline{q}_{\infty
}\left( x\right) dx=\frac{1}{2t}\mathbf{E}_{\overline{q}_{\infty }}\left[
\mu _{t}\left( \phi \right) ^{2}-\mu _{t}\left( \phi ^{2}\right) \right]
\end{equation*}
such that $t\left[ 1-\mathbf{P}\left( N_{t}\left( x\right) =0\right) \right] 
\underset{t\rightarrow \infty }{\rightarrow }\mu ^{-1}\overline{\phi }%
_{\infty }\left( x\right) ,$ uniformly in $x.$

$\left( iii\right) $ For all bounded measurable function $\psi $ on $I:$ 
\begin{equation*}
\frac{1}{t}\mathbf{E}\left[ \sum_{n=1}^{N_{t}\left( x\right) }\psi \left( 
\widetilde{x}_{t}^{\left( n\right) }\right) \mid N_{t}\left( x\right)
>0\right] \underset{t\rightarrow \infty }{\rightarrow }\mu \int_{\left(
0,1\right) }\psi \left( y\right) \overline{q}_{\infty }\left( y\right) dy.
\end{equation*}
\newline

From $\left( i\right) $, it is clear that the process gets ultimately
extinct with probability $1.$ In the trade-off between killing-branching and
reflection at the boundaries, all particles get eventually absorbed but the
global BD process turns out be critical. Thus, the killing part of $\lambda
\left( x\right) $ is strong enough to avoid the explosion of the number of
particles inside the unit interval, resulting in an overall critical process
where global extinction still holds.

In the statement $\left( ii\right) ,$ $1-\mathbf{P}\left( N_{t}\left(
x\right) =0\right) =\mathbf{P}\left( T\left( x\right) >t\right) $ where $%
T\left( x\right) $ is the global extinction time of the particle system. The
Pareto tails of $T\left( x\right) $ decay like $t^{-1}$, thus algebraically
slowly: the time till extinction in this critical model is much longer than
in the previous neutral sub-critical case (with exponential tails). From $%
\left( ii\right) ,$ $\overline{\phi }_{\infty }\left( x\right) $ has again a
natural interpretation in terms of the propensity of the particle system to
survive to its extinction fate.

$\left( iii\right) $ with $\psi =1$ reads $\frac{1}{t}\mathbf{E}\left[
N_{t}\left( x\right) \mid N_{t}\left( x\right) >0\right] \underset{%
t\rightarrow \infty }{\rightarrow }\mu $ giving an interpretation of the
constant $\mu $. The constant $\mu $ is also (\cite{AH2}, page $287$) 
\begin{equation*}
\mu =\frac{1}{2}\lambda _{*}\int_{0}^{1}\kappa \left( x\right) \overline{%
\phi }_{\infty }\left( x\right) ^{2}\overline{q}_{\infty }\left( x\right)
dx=\lambda _{*}\int_{0}^{1}p_{2}\left( x\right) \overline{\phi }_{\infty
}\left( x\right) ^{2}\overline{q}_{\infty }\left( x\right) dx<\infty
\end{equation*}
and so is explicitly available in our case.\newline

The ground states of $\overline{G}+\lambda _{0}$ and its adjoint are thus $%
\left( \overline{\phi }_{\infty },\overline{q}_{\infty }\right) $ and
explicit here. It is also useful to consider the process whose infinitesimal
generator is given by the Doob-transform 
\begin{equation*}
\overline{\phi }_{\infty }^{-1}\overline{G}\left( \overline{\phi }_{\infty
}\cdot \right) =\overline{\phi }_{\infty }^{-1}\left( \widetilde{G}+\lambda
\right) \left( \overline{\phi }_{\infty }\cdot \right) ,
\end{equation*}
because product-criticality is preserved under this transformation. The
ground states associated to this new operator and its dual are $\left( 1,%
\overline{\phi }_{\infty }\overline{q}_{\infty }\right) $. Developing, we
obtain a process whose infinitesimal generator is 
\begin{equation*}
\widetilde{G}+\frac{\overline{\phi }_{\infty }^{\prime }}{\overline{\phi }%
_{\infty }}g^{2}\partial _{x}=G+\frac{u_{0}^{\prime }}{u_{0}}g^{2}\partial
_{x}=G,
\end{equation*}
with no multiplicative part. The associated diffusion process is the
starting point WF diffusion with mutations, which is positive recurrent and
so its invariant measure $\overline{\phi }_{\infty }\overline{q}_{\infty
}=u_{0}v_{0}=m\left( y\right) $ is integrable. \newline

\textbf{Remark.} The functional generating function $u\left( x,t;z\right) =%
\mathbf{E}\left[ \prod_{n=1}^{N_{t}\left( x\right) }z^{\psi \left( 
\widetilde{x}_{t}^{\left( n\right) }\right) }\right] $ of the measure-valued
branching particle system obeys now the nonlinear (quadratic) PDE: 
\begin{equation*}
\partial _{t}u\left( x,t;z\right) =\lambda _{*}\theta \left( x,u\left(
x,t;z\right) \right) +\widetilde{G}\left( u\left( x,t;z\right) \right) ;%
\text{ }u\left( x,0;z\right) =z^{\psi \left( x\right) },
\end{equation*}
where $\theta \left( x,z\right) =\mathbf{E}\left[ z^{M\left( x\right)
}\right] -z=\left( p_{2}\left( x\right) z^{2}+p_{0}\left( x\right) \right)
-z $ or 
\begin{equation*}
\theta \left( x,z\right) =\left( z-1\right) \left( p_{2}\left( x\right)
z-p_{0}\left( x\right) \right)
\end{equation*}
is the shifted probability generating function of the branching law of $%
M\left( x\right) .$

If $u\left( x,t\right) :=\partial _{z}u\left( x,t;z\right) _{z=1}=\mathbf{E}%
\left[ \sum_{n=1}^{N_{t}\left( x\right) }\psi \left( \widetilde{x}%
_{t}^{\left( n\right) }\right) \right] $, recalling $\lambda \left( x\right)
=\lambda _{*}\left( p_{2}\left( x\right) -p_{0}\left( x\right) \right) ,$ $%
u\left( x,t\right) $ obeys the linear backward PDE 
\begin{equation*}
\partial _{t}u\left( x,t\right) =\lambda \left( x\right) u\left( x,t\right) +%
\widetilde{G}\left( u\left( x,t\right) \right) ;\text{ }u\left( x,0\right)
=\psi \left( x\right)
\end{equation*}
involving $\overline{G}\left( \cdot \right) =\widetilde{G}\left( \cdot
\right) +\lambda \left( x\right) \cdot $. It holds that 
\begin{equation*}
u\left( x,t\right) =\mathbf{E}_{x}\left( e^{\int_{0}^{t}\lambda \left( 
\widetilde{x}_{s}\right) ds}\psi \left( \widetilde{x}_{t}\right) \right) .%
\text{ }\diamond
\end{equation*}

\end{document}